\documentclass[prb,twocolumn,amssymb,amsmath,aps,showpacs,floatfix,makecell]{revtex4-2}
\usepackage[colorlinks=true,linkcolor=blue,citecolor=blue,urlcolor=blue]{hyperref}
\usepackage{graphicx,bm}
\usepackage{xcolor}
\begin{document}

\title{Acoustic plasmons and isotropic short-range interaction\\ in two-component electron liquids}

%\date{\today}

\author{A.~N.~Afanasiev}
\email{afanasiev.an@mail.ru}
\affiliation{Ioffe Institute, St.~Petersburg 194021, Russia}

\begin{abstract}
Dispersion of acoustic plasmons and isotropic Landau parameters are calculated in three- and two-dimensional two-component electron-electron and electron-hole liquids at various concentration and mass ratios using Landau-Silin kinetic equation and the random phase approximation for the self-energy. It is shown that the mode propagation and the strength of quasiparticle interaction are determined by the intercomponent screening and are asymmetric with respect to charge composition of two-component liquid. The well-defined acoustic plasmon-zero sound mode arises at strong difference in concentrations between components and its renormalization by the short-range exchange-correlation interaction is negligible. The acoustic plasmon mediated interparticle effective interaction in the fast component is weak in both the three- and two-dimensional electron liquids with parabolic dispersion, so the associated plasmonic superconductivity and the formation of acoustic plasmarons are unfavorable.
\end{abstract}

\maketitle
  
%%%%%%%%%%%%%%%%%%%%%%%%%%%%%%%%%%%%%%%%%%%%%%%%%%%%%%%%%%%%%%%%%%%%%%%%%%%%%%%%%%%%%%%%%%%%%%%%%%%%%%%%%%%
%%%%%%%%%%%%%%%%%%%%%%%%%%%%%%%%%%%%%%%%%%%%%%%%%%%%%%%%%%%%%%%%%%%%%%%%%%%%%%%%%%%%%%%%%%%%%%%%%%%%%%%%%%%

\section{Introduction}

The concept of elementary excitations, originally introduced by L.~D.~Landau~\cite{Landau_He_II,Landau_FL:QP,Landau_FL:ZS,Pines_QL} to describe the low-energy properties of liquid helium, plays a key role in condensed matter physics. In particular, collective modes which represent the bosonic part of the excitation spectra are well-known to be responsible for various electronic and structural properties of condensed matter systems~\cite{Rodin2020,Pines_EES_2018}.

The recent demonstration of the strongly interacting electron liquid (EL) and Dirac fluid behavior in high-quality samples of graphene~\cite{Polini2020,Lucas2018_Review}, quasi-two dimensional metals~\cite{Moll2016} and GaAs quantum wells~\cite{Gusev2018_AIP} manifesting in the hydrodynamic regime of electron transport has stimulated an extensive research on the collective mode structure of such systems. Novel hydrodynamic and collisionless electronic sound modes were predicted in Dirac fluids~\cite{Svintsov2018,Kiselev2020,Narozhny2021} and gated two-dimensional systems~\cite{Torre2019}. However, the revisited spectrum of collective excitations of ELs was shown to be consistent with the fundamental result of Landau-Silin theory~\cite{Silin1958,Silin1959}: in the long-wavelength domain the long-range coulomb interaction transforms both the collisionless zero sound~\cite{Silin1958} and the hydrodynamic first sound~\cite{Lucas2018,Jian2021} into plasmons~\cite{BohmPines1952}. Sound modes supported by EL are associated with the anisotropic part of the short-range quasiparticle interaction described by higher-order Landau parameters $F_n$, $n\geq 1$~\cite{Silin1959,Gorkov1963,Klein2019,Aquino2019,Aquino2020}. Namely, propagation of the shear sound~\cite{Alekseev2018,Alekseev2019_SC,Alekseev2019_PRL,Khoo2019,Zhang2021} determined by the dipole Landau interaction manifests in the the viscoelastic resonance in highly viscous electron fluids. Absence of density oscillation in this mode provides immunity to coulomb interaction, but also makes it inaccessible by standard charge-sensitive experimental techniques~\cite{Khoo2020,Valentinis2021_SciRep,Valentinis2021_PRR}.

The spectrum of collective density oscillations of two-component electron systems is qualitatively different. In high-density two-component degenerate electron gas, the ordinary high-frequency {\it optical} plasma mode is accompanied by {\it acoustic plasmon} in quasi-classical domain, see Fig.~\ref{Fig:Scheme}. The character of density oscillations in this mode in long wavelength limit is similar to zero sound in neutral Fermi liquids~\cite{Landau_FL:ZS, Landau_V9}: the total charge variation in the vibration is zero since the partial contributions of components compensate each other (Fig.~\ref{Fig:Scheme}). The mode has gapless linear dispersion and its velocity is between the Fermi velocities of the slow and the fast components. Predicted long ago by D.~Pines~\cite{Pines1956}, acoustic plasmons (APs) were considered to be a source of unconventional, purely electronic superconductivity in multiband materials such as transition metals with incomplete inner shell~\cite{Frohlich1968}, electron-hole liquids~\cite{Vignale1985,Vignale1989}, graphene/transition metal dichalcogenide bilayers~\cite{Fatemi2018}, bismuth~\cite{Ruhman2017} and layered materials such as twisted bilayer graphene~\cite{Sharma2020}. Another manifestation of this mode is the spin plasmon in two-dimensional degenerate gases with spin splitting~\cite{Agarwal2014,Kreil2015,Xiao2017,Enaldiev2018,Schober2020}. At low densities, modification of AP dispersion in two-component ELs was studied within the generalized random phase approximation in Refs.~\cite{Vignale1982,Vignale1988}. Undamped zero sound mode separated from the particle-hole continuum of two-component charged Fermi liquids was considered in the framework of Landau-Silin theory~\cite{Dunin1972,Akhiezer1974,Oliva1982} and no connection between these two modes has been established yet.

In this work we study collisionless sound modes of three- and two-dimensional two-component ELs with the microscopically calculated isotropic short-range interaction. We show that the upper undamped branch is absent and zero sound in two-component ELs is represented solely by the AP mode. Renormalization of the well-defined AP mode by the isotropic exchange-correlation interaction is negligible, however, the latter introduces non-equivalence of APs in electron-electron and electron-hole liquids. AP propagation regime characterized by the strong AP mediated Fr\"ohlich-like interaction necessary for plasmonic superconductivity and formation of acoustic plasmarons is suppressed in 3D and hardly achievable in 2D ELs with parabolic dispersion.

The work is structured as follows. In Section~\ref{Sec:AP_Landau_Silin} we formulate the system of macroscopic collisionless Landau-Silin kinetic equations for two-component ELs, establish identity between AP and zero sound, and obtain dispersion equation for AP. In Section~\ref{Sec:AP_Structure} we study regimes of AP propagation and consider AP velocity and damping within the random phase approximation (RPA). Next, in Section~\ref{Sec:LF} we calculate isotropic Landau parameters of the two-component EL determined by the short-range exchange-correlation interaction between quasiparticles. In Section~\ref{Sec:Renorm_AP} we apply the microscopically calculated Landau parameters to macroscopic theory of Section~\ref{Sec:AP_Landau_Silin} to determine the renormalization of AP velocity and damping in electron-electron and electron-hole liquids. The strength of the AP mediated electron-electron interaction in the fast component of two-component ELs and the case of type-I Weyl semimetals are discussed in Section~\ref{Sec:Discussion}. Appendix contains explicit forms of isotropic Ladau parameters suitable for further numerical calculation.

%%%%%%%%%%%%%%%%%%%%%%%%%%%%%%%%%%%%%%%%%%%%%%%%%%%%%%%%%%%%%%%%%%%%%%%%%%%%%%%%%%%%%%%%%%%%%%%%%%%%%%%%%%%
%%%%%%%%%%%%%%%%%%%%%%%%%%%%%%%%%%%%%%%%%%%%%%%%%%%%%%%%%%%%%%%%%%%%%%%%%%%%%%%%%%%%%%%%%%%%%%%%%%%%%%%%%%%
\section{Equivalence of acoustic plasmon and zero sound}
\label{Sec:AP_Landau_Silin}

We consider isotropic, non-polarized system with two types of degenerate carriers with parabolic dispersion. Each component is characterized by elementary charge $e_{\alpha}=\pm e$, mass $m_{\alpha}$ and Fermi wave vector $k_{\alpha}$. Hereinafter, indices $\alpha=1,2$ and $\beta=1,2$ denote the components of EL. The evolution of the nonequilibrium quasiparticle distributions in components associated with propagation of AP is described by the pair of coupled collisionless Landau-Silin kinetic equations~\cite{Landau_FL:ZS,Silin1959,Landau_V9} of the form
\begin{equation}
	\label{Eq:KinEq}
	\frac{\partial n_{\alpha}}{\partial t} + \frac{\partial \mathcal{E}_{\alpha}}{\partial {\bf p}} \frac{\partial n_{\alpha}}{\partial {\bf r}} - \frac{\partial \mathcal{E}_{\alpha}}{\partial {\bf r}}\frac{\partial n_{\alpha}}{\partial {\bf p}} =0
\end{equation}
where $n_{\alpha}({\bf r},{\bf p}, t)$ is the nonequlibrium occupation number and the quasiparticle energies $\mathcal{E}_{\alpha}({\bf r},{\bf p}, t)=\mathcal{E}_{\alpha}[n_1,n_2]$ are functionals of $n_{\alpha}({\bf r}, {\bf p},t)$ given by
\begin{gather}
	\label{Eq:EnergyRenorm}
	\mathcal{E}_\alpha({\bf r}, {\bf p}, t)=\epsilon_{\alpha}({\bf p})+ U_{\alpha}({\bf r}, {\bf p}, t) + E_{\alpha}(\textbf{r},\textbf{p},t) ,\\
	\label{Eq:SelfConsPotential}
	U_{\alpha}({\bf r}, {\bf p}, t)=\sum\limits_{\beta\textbf{r}'} \frac{e_{\alpha} e_{\beta}}{\varkappa |{\bf r}-{\bf r}'|}n_{\alpha}({\bf r},{\bf p}, t) \delta N_{\beta}({\bf r}', t) ,\\
	\label{Eq:QP_energy}
	E_{\alpha}(\textbf{r},\textbf{p},t)=\sum\limits_{\beta \textbf{p}'}f^{\alpha\beta}_{s}(\vartheta)\delta n_{\beta}(\textbf{r},\textbf{p}',t) ,
\end{gather}
where $\epsilon_{\alpha}(\textbf{p})$ is the bare energy of non-interacting fermions, $U_{\alpha}(\textbf{r},\textbf{p},t)$ is the potential energy in the self-consistent electrostatic field and $E_{\alpha}(\textbf{r},\textbf{p},t)$ is the energy renormalization by the spin-symmetric part of short-range Landau interaction between quasiparticles, $\delta N_{\alpha}({\bf r}',t)$ stands for the nonequilibrium concentration, $\varkappa$ is the background dielectric constant and $\sum_{\textbf{r}'}=\int d {\bf r}' $. In isotropic systems, Landau interaction function depends only on the angle $\vartheta=\widehat{\textbf{p}\textbf{p}'}$ between momenta of interacting quasiparticles. In two-component ELs the diagonal elements of $f_s^{\alpha\beta}(\vartheta)$ describe the intracomponent short-range interaction at the Fermi surfaces, while the non-diagonal ones describe intercomponent interaction.

Following Landau-Silin theory of EL~\cite{Landau_FL:ZS,Silin1959,Landau_V9}, we describe the weakly excited state of two-component EL in terms of the deformed Fermi surfaces $\mathcal{E}_F^{(\alpha)}({\bf r}, {\bf p}, t)=\mu_{\alpha}+\Phi_{\alpha}({\bf r}, {\bf p}, t)$ in components, where $\mu_{\alpha}$ is the equilibrium Fermi energy and $\Phi_{\alpha}({\bf r},{\bf p},t)\ll \mu_{\alpha}$ is the Fermi surface deformation. Hence the occupation numbers take the form
\begin{equation}
	\label{Eq:NonEqDistr}
	n_{\alpha}({\bf r}, {\bf p}, t)\approx n_0^{(\alpha)}({\bf p}) + \Phi_{\alpha}({\bf r}, {\bf p}, t)\delta\left(\mu_{\alpha}-\epsilon_{\alpha}({\bf p})\right) ,
\end{equation}
where $n_{0}^{(\alpha)}(\textbf{p})$ is the equilibrium distribution. Therefore, kinetic equations~(\ref{Eq:KinEq}) can be linearized and reduced to equations for the Fermi surface deformation. According to~(\ref{Eq:NonEqDistr}), quasiparticles are excited only from the Fermi surface and the deformation $\Phi_{\alpha}({\bf r}, {\bf p}_{\alpha}, t)=\Phi_{\alpha}({\bf r},{\bf e_p}, t)$ depends only on the direction of Fermi velocity ${\bf v}_{\alpha}=\nabla_{\bf p}\epsilon_{\alpha}( \textbf{p}_{\alpha})$, where $p_{\alpha}=\hbar k_{\alpha}$ denotes the Fermi momentum.

%%%%%%%%%%%%%%%%%%%%%%%%%
\begin{figure}[t!]
	\includegraphics[]{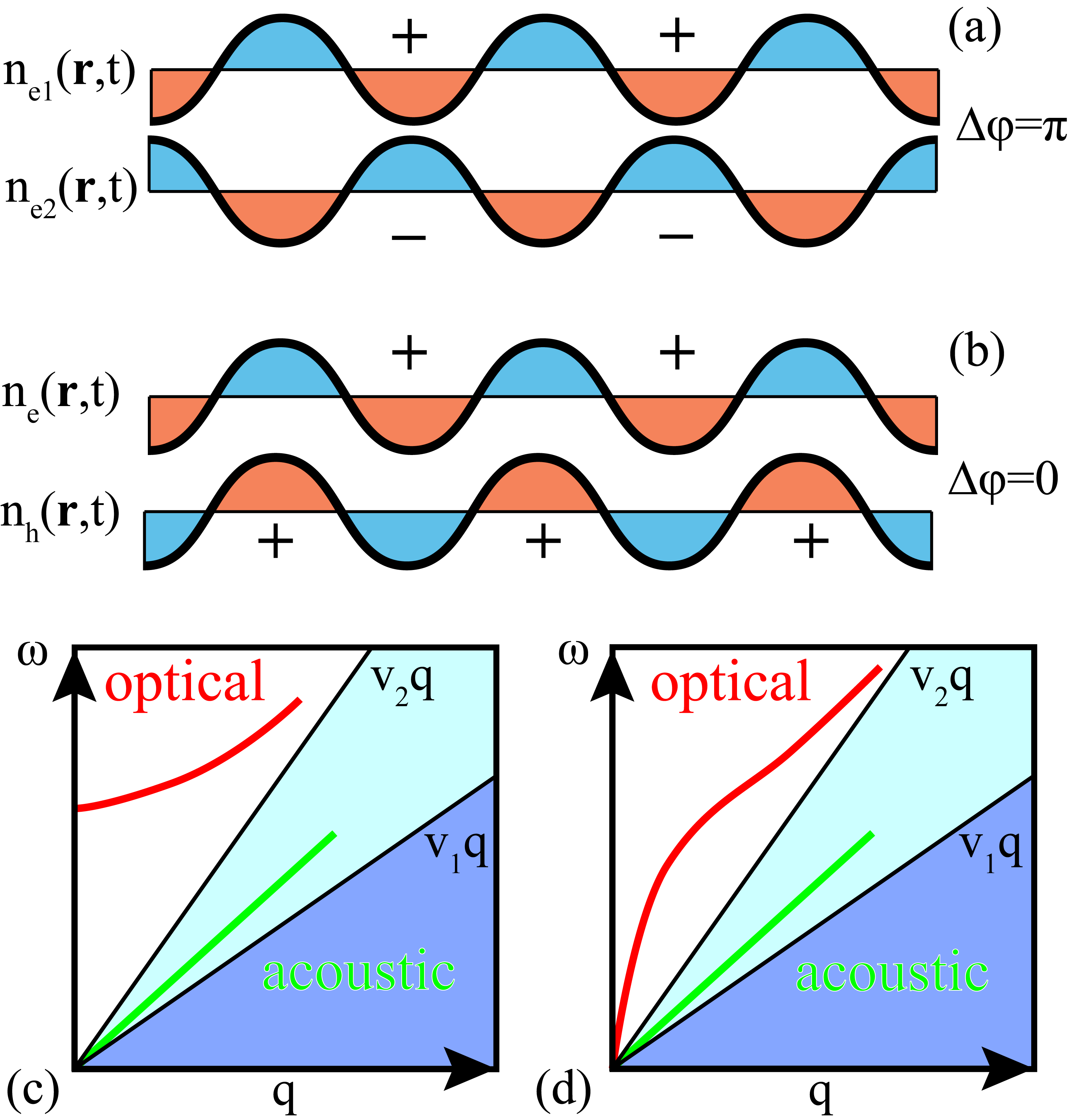}
	\caption{Schematic representation of acoustic plasmon mode in two-component electron liquids. Panels (a) and (b) demonstrate the characteristic density (black solid lines) and charge (marked by color) variations in the vibration in electron-electron and electron-hole liquid, respectively. Panels (c) and (d) demonstrate schematically plasmon dispersions and particle-hole continuum domains (filled by color) in three and two dimensions.}
	\label{Fig:Scheme}
\end{figure}
%%%%%%%%%%%%%%%%%%%%%%%%%

Microscopic calculations show that the spin-symmetric part of Landau interaction function in the single-component ELs is almost isotropic~\cite{Pines_Nozieres_2018,Giuliani_Vignale_2005}. Namely, dimensionless isotropic Landau parameter $F_0$ is greater than the dipole parameter $F_1$ and the other higher-order ones by an order of magnitude in 3D and 2D. In this work we consider the effect of isotropic part of the short-range quasiparticle interaction on AP propagation in two-component ELs. Consequently, Landau interaction function in~(\ref{Eq:QP_energy}) is approximated by $f_s^{\alpha\beta}(\vartheta)\approx D_{\beta}^{-1} F_0^{\alpha\beta}$ and we neglect the interaction-induced renormalization of mass and the density of states at the Fermi surfaces $D_{\beta}=D_{d}(\mu_{\beta})$. Emergence of transverse zero sound in two-component ELs associated with $F_1^{\alpha\beta}$ was considered in~\cite{Romanov2008}.

In this work we consider the long wavelength domain, where AP has linear dispersion $\omega_{\rm ac}(k)=sk$ characterized by the velocity of sound $s$. Propagation of the mode is accompanied by the plane wave-like deformation of the Fermi surfaces in components
\begin{equation}
	\label{Eq:AcPlFermiDeform}
	\Phi_{\alpha}({\bf r},{\bf e_p},t)=\Phi_s^{(\alpha)}(\theta) {\rm e}^{i({\bf kr}-skt)} ,
\end{equation}
where $\theta$ is the angle between ${\bf e_p}$ and the direction of the wave propagation ${\bf k}/k$. Combining Eqs.~(\ref{Eq:NonEqDistr}) and~(\ref{Eq:AcPlFermiDeform}) with~(\ref{Eq:KinEq}),~(\ref{Eq:SelfConsPotential}) and~(\ref{Eq:QP_energy}), we come to important result that the propagation of AP, i.e. its velocity and the shape of the oscillating Fermi surfaces, is governed by the following system of two coupled integral equations
\begin{equation}
	\label{Eq:AcPl_FLsystem}
	\left(\frac{s}{v_{\alpha}}-\cos\theta\right)\Phi^{(\alpha)}_{s}(\theta)=\cos\theta \int \frac{d \Omega'}{\Omega_d} F^{\alpha\beta}(k) \Phi_{s}^{(\beta)}(\theta') ,
\end{equation}
where $\Omega_d=\int d\Omega'$ is the $d$-dimensional solid angle and the summation over repeated indices is assumed. System~(\ref{Eq:AcPl_FLsystem}) reproduces the form of zero sound equations in two-component Fermi liquids~\cite{Dunin1972,Akhiezer1974,Oliva1982} with the isotropic $k$-dependent Landau parameters renormalized~\cite{Silin1959,Lucas2018,Jian2021} by the long-range coulomb interaction
\begin{gather}
	\label{Frenorm}
	F^{\alpha\beta}(k)=F^{\alpha\beta}_0 + F^{\alpha\beta}_{c}(k) ,\\
	F^{\alpha\beta}_{c}(k)=D_{\beta}V_{\alpha\beta}(k)=\frac{e_{\alpha}}{e_{\beta}}\left(\frac{\kappa_{\beta}}{k}\right)^{d-1} ,
\end{gather}
where $\kappa_{\alpha}$ denote the Thomas-Fermi wave vectors in components and the Fourier transform of coulomb interaction in 3D and 2D is $V_{\alpha\beta}(k)=4\pi e_{\alpha}e_{\beta}/\varkappa k^2$ and $V_{\alpha\beta}(k)=2\pi e_{\alpha}e_{\beta}/\varkappa k$, respectively. Explicit expressions for the complex amplitudes of the Fermi surface deformations following from~(\ref{Eq:AcPl_FLsystem}) are
\begin{equation}
	\label{Eq:AcPl_F}
	\Phi_s^{(\alpha)}(\theta)=\mathcal{U}_s^{(\alpha)}\frac{\cos\theta}{s/v_{\alpha}-\cos\theta} ,
\end{equation}
where $\mathcal{U}_s^{(\alpha)}$ denote the complex amplitudes of the total self-consistent potential in components.

When the particular form of $\Phi_s^{(\alpha)}(\theta)$ given by Eq.~(\ref{Eq:AcPl_F}) is inserted into~(\ref{Eq:AcPl_FLsystem}) and the angular integration is performed, the system of integral equations for $\Phi_s^{(\alpha)}$ is reduced to homogeneous system of linear equations for the complex amplitudes of the self-consistent potential
\begin{equation}
	\label{Eq:Syst_U}
	[\delta_{\alpha\beta}-F^{\alpha\beta}(k)\tilde{\Pi}_{\beta}(s)]\mathcal{U}_{s}^{(\beta)}=0 ,
\end{equation}
where
\begin{equation}
	\label{Eq:P}
	\tilde{\Pi}_{\alpha}(s)=\int\frac{d \Omega}{\Omega_d}\frac{\cos \theta}{s/v_{\alpha}-\cos \theta + i0} ,
\end{equation}
is the dimensionless non-interacting quasi-classical polarizability. Non-trivial solution of~(\ref{Eq:Syst_U}) exists when $|\delta_{\alpha\beta}-F^{\alpha\beta}(k)\tilde{\Pi}_{\beta}(s)|=0$. In the long wavelength limit this equation reduces to equation for the AP velocity 
\begin{equation}
	\label{Eq:AcPl_Disp}
	F^{11}_{c}(k) \overline{\Pi}_1(s)+F^{22}_{c}(k) \overline{\Pi}_2(s)=0 ,
\end{equation}
which can be formulated in terms of the dimensionless proper polarizabilities
\begin{equation}
	\label{Eq:Proper_P}
	\overline{\Pi}_{\alpha}(s)=\frac{\tilde{\Pi}_{\alpha}(s)}{1-\tilde{F}_0^{\alpha\alpha}\tilde{\Pi}_{\alpha}(s)} ,
\end{equation}
determined by the reduced Landau parameters
\begin{equation}
	\label{Eq:Reduced_F}
	\tilde{F}_0^{\alpha\alpha}=F_0^{\alpha\alpha}-\frac{e_{\alpha}}{e_{\beta}}F_0^{\beta\alpha},\, \alpha\neq\beta .
\end{equation}
The dispersion equation~(\ref{Eq:AcPl_Disp}) has two possible solutions. First of them lies beyond the particle-hole continuums of components and its velocity is greater than both the velocities of the slow (first) and the fast (second) carriers. This is a conventional zero sound mode~\cite{Dunin1972,Akhiezer1974,Oliva1982} analogous to the case of single-component neutral Fermi liquid. However, this branch arises only when the reduced short-range interaction $\tilde{F}_0^{\alpha\alpha}$ is repulsive in both components. As it will be shown in Section~\ref{Sec:LF}, this is not the case of two-component ELs, thus this mode is absent. Velocity of the second branch lies between the slow and the fast Fermi velocities. This is an AP mode, which is a special case of zero sound in two-component ELs. Acoustic plasmon velocity determined by Eq.~(\ref{Eq:AcPl_Disp}) coincides with the one given by the zero of the quasi-classical dielectric function of two-component EL in the long wavelength limit
\begin{gather}
	\label{Eq:DielectricFunction}
	\varepsilon(sk,k)=1-V(k)\bar{\chi}(s) ,\\
	\label{Eq:Proper_density_response}
	\bar{\chi}(s)=\frac{D_1 \tilde{\Pi}_1[1-\tilde{F}_0^{22}\tilde{\Pi}_2]+D_2 \tilde{\Pi}_2[1-\tilde{F}_0^{11}\tilde{\Pi}_1]}{\Delta} ,\\
	\begin{split}
		\Delta(s)=1-F_0^{11}\tilde{\Pi}_1-F_0^{22}\tilde{\Pi}_2+\\
			+(F_0^{11}F_0^{22}-F_0^{12}F_0^{21})\tilde{\Pi}_1\tilde{\Pi}_2 .
	\end{split}
\end{gather}
Here $V(k)=V_{\alpha\alpha}(k)$ and $\bar{\chi}(s)$ denotes the proper density response function. Eq.~(\ref{Eq:AcPl_Disp}) is equivalent to the vanishing proper density response $\bar{\chi}(s)=0$. 

According to Eqs.~(\ref{Eq:Syst_U}), the phase shift between the complex amplitudes of the density oscillations in components $\delta N_{s}^{(\alpha)}=D_{\alpha}\tilde{\Pi}_s^{(\alpha)}\mathcal{U}_s^{(\alpha)}$ in the long-wavelength limit dictated by the long-range intra- and intercomponent coulomb interaction depends on the charge composition of EL
\begin{equation}
	\label{Eq:Phase_relation}
	\delta N_s^{(1)}=-\frac{e_2}{e_1}\delta N_s^{(2)} .
\end{equation}
Density oscillations are out-of-phase in the case of electron-electron liquid and in-phase in electron-hole liquid, see Fig.~\ref{Fig:Scheme}a,b. Therefore, the mode carries zero total charge density $\delta\rho_s=e_1\delta N_s^{(1)}+e_2 \delta N_s^{(2)}=0$. 

Since AP velocity is between the slow and the fast Fermi velocities, the mode is affected by Landau damping~\cite{Landau_V10} due to the intraband single particle transitions in the fast component. Therefore AP frequency acquires imaginary part and we treat it in terms of the complex velocity $s=s'-i s''$, where the correct sign of $s''>0$ corresponds to decay of the charge density oscillations in time. Well-defined AP with $s''/s' \ll 1$ arises when the difference in the Fermi velocities is great $v_2/v_1 \gg 1$ implying that the liquid components are weakly coupled. Taking into account an approximate form of the polarizability ${\rm Re}\,\tilde{\Pi}_{2}(s')=-1$ at $s'/v_2 \rightarrow 0$, the dispersion equation~(\ref{Eq:AcPl_Disp}) transforms into
\begin{equation}
	\label{Eq:AcPl_ZS}
	\tilde{\Pi}_1(s')=\frac{1}{F_{\rm eff}} ,
\end{equation}
indicating that in this regime AP is identical to the zero sound~\cite{Landau_V9} propagating in the slow component with the effective interaction parameter
\begin{equation}
	\label{Eq:Feff}
	F_{\rm eff}=(1+\tilde{F}_0^{22})F_{_{\rm RPA}}+\tilde{F}_0^{11} .
\end{equation}
Note that $\tilde{\Pi}_1(s')$ is real-valued function at $s'>v_1$. This special case of zero sound does not vanish in the high-density regime, when the short range interaction is absent $F_0^{\alpha\beta}\to 0$ and Eqs.~(\ref{Eq:AcPl_Disp}),~(\ref{Eq:DielectricFunction}) and~(\ref{Eq:Proper_density_response}) reproduce the results for the quasi-classical limit of RPA. The effective interaction parameter~(\ref{Eq:Feff}) in RPA is determined by the coulomb interaction in the slow component at the Thomas-Fermi wave vector of the fast component $F_{_{\rm RPA}}=F_{c}^{11}(\kappa_2)=(\kappa_1/\kappa_2)^{d-1}$. This reveals the reason why the acoustic plasmon-zero sound mode is stable to the long-range coulomb interaction: particles of the fast component screen out the intracomponent interaction in the slow one, making it effectively short-range. However, the mode stabilization inevitably leads to the finite lifetime of this collisionless zero sound mode introduced by the particle-hole excitations in the fast component
\begin{equation}
	\label{Eq:AP_damping}
	s''\approx \frac{F_{_{\rm RPA}}}{F^2_{\rm eff}}\frac{{\rm Im}\, \tilde{\Pi}_2(s')}{\partial {\rm Re}\,\tilde{\Pi}_1(s') /\partial s} .
\end{equation}

%%%%%%%%%%%%%%%%%%%%%%%%%%%%%%%%%%%%%%%%%%%%%%%%%%%%%%%%%%%%%%%%%%%%%%%%%%%%%%%%%%%%%%%%%%%%%%%%%%%%%%%%%%%
%%%%%%%%%%%%%%%%%%%%%%%%%%%%%%%%%%%%%%%%%%%%%%%%%%%%%%%%%%%%%%%%%%%%%%%%%%%%%%%%%%%%%%%%%%%%%%%%%%%%%%%%%%%

\section{Structure of the acoustic plasmon mode}
\label{Sec:AP_Structure}
 
%%%%%%%%%%%%%%%%%%%%%%
\subsection{Shape of the Fermi surfaces in acoustic plasmon mode}
\label{Sec:Shape}

%%%%%%%%%%%%%%%%%%%%%%%%%
\begin{figure}[t!]
	\includegraphics[width=\linewidth]{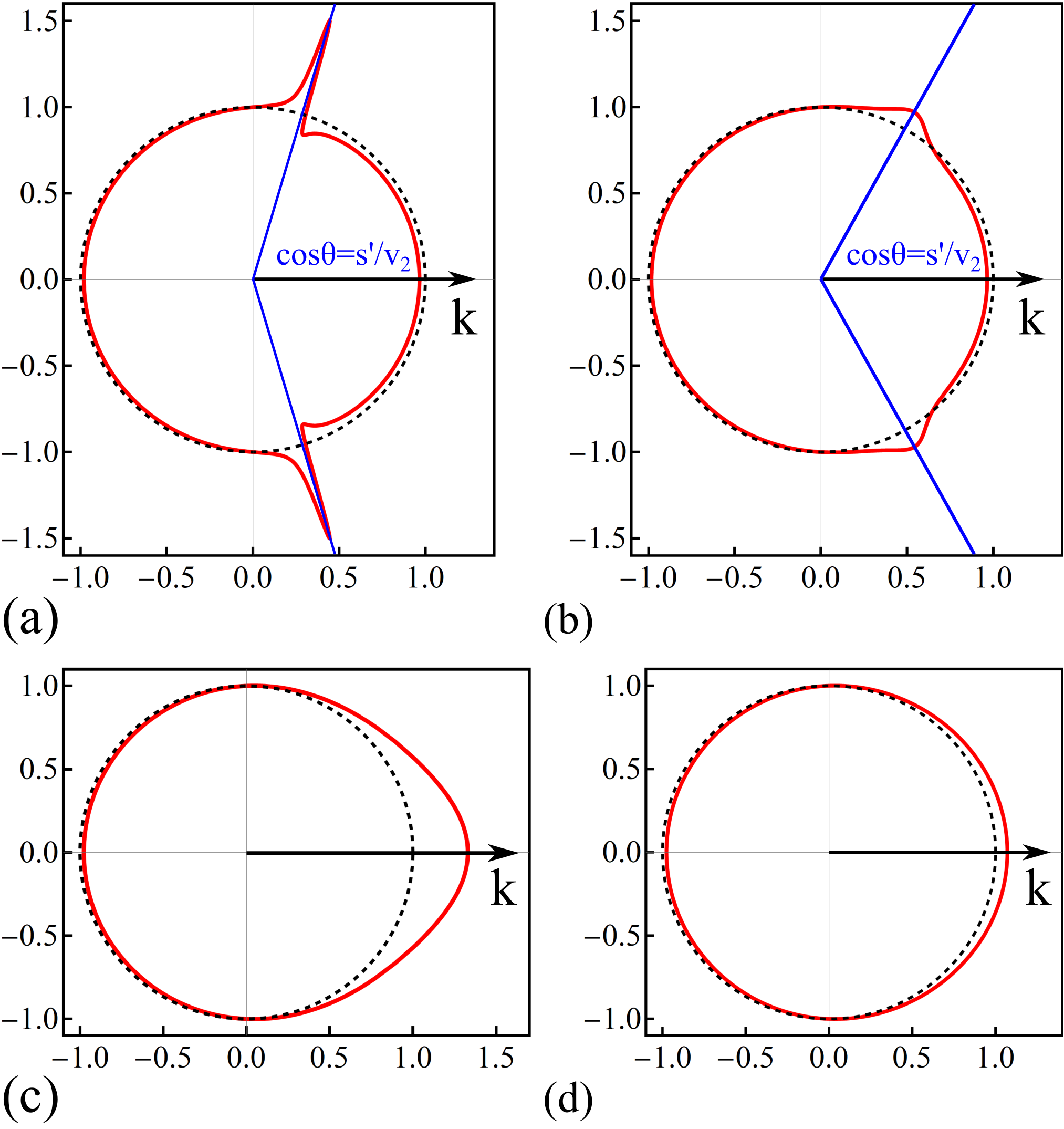}
	\caption{Shapes of the Fermi surfaces in the fast (a,b) and the slow (c,d) components of electron-electron liquid corresponding to acoustic plasmon propagation in (a,c) the regime of zero sound in the almost ideal Fermi gas described by Eqs.~(\ref{Eq:Weak_3D}) and~(\ref{Eq:Weak_2D}) at $s'/v_1 =1.1$ and (b,d) the ion-acoustic wave regime after Eqs.~(\ref{Eq:Strong_3D}) and~(\ref{Eq:Strong_2D}) at $s'/v_1=2$. Red curves denote the cross-section of the deformed Fermi surfaces $\mathcal{E}_F^{(1,2)}({\bf r},{\bf p}, t)/\mu_{1,2}$ as a function of ${\bf e_p}$ with respect to plasmon wave vector ${\bf k}$ and the dashed black circles correspond to the equilibrium case. The ${\bf kr}-\omega t=7\pi/4$ wave plane is chosen and ${\rm Re}\,\mathcal{U}^{(1,2)}_s/\mu_2 = 0.035$, $\mu_2/\mu_1=2$, $v_2/v_1=4$, $s''/v_2 = 0.05$.}
	\label{Fig:Shapes}
\end{figure}
%%%%%%%%%%%%%%%%%%%%%%%%%

Now we turn to investigation of the internal structure of AP and consider the shape of the Fermi surfaces which oscillate as the mode propagates. Since the AP velocity has both real and imaginary parts, we use real representations of $\Phi_{\alpha}({\bf r},\theta,t)={\rm Re}\left(\Phi_s^{(\alpha)}(\theta) {\rm e}^{-s'' k t} {\rm e}^{i({\bf kr}-s'kt)}\right)$ to study the angular dependence of the Fermi surfaces. Landau damping (i.e. $s''\neq 0$) not only provides the exponential decay of $\Phi_{1,2}$ but also sufficiently affects the shape of the Fermi surface in the fast component, as it is shown in Fig.~\ref{Fig:Shapes}. Nonzero $s''$ regularize singularities in $\Phi_2$ at $\cos\theta=s'/v_2 $ signifying excitation of quasiparticles of the fast component which move in phase with the wave. This resonant condition correspond to the inverse ${\rm \check{C}}$herenkov effect~\cite{Landau_V10} which underlies Landau damping. The resulting shape of $\mathcal{E}_F^{(2)}({\bf r},\theta,t)$ resemble the ${\rm \check{C}}$herenkov cone indicating the propagation of the shock wave of the fast Fermi surface deformation in the energy space.

In the case of the slow component, ${\rm \check{C}}$herenkov pole is absent since $s'\geq v_1$ and the influence of Landau damping on the shape of $\Phi_s^{(1)}(\theta)$ is negligible when $s''/s' \ll 1$. Therefore the shape of the slow Fermi surface given by~\cite{Us_Note}
\begin{equation}
	\label{Eq:ZeroSoundWave}
	\Phi_1({\bf r},\vartheta,t)=\mathcal{U}^{(1)}_s {\rm e}^{-s'' k t} \frac{\cos\theta}{s'/v_1-\cos\theta}\cos({\bf kr}-s'kt) ,
\end{equation}
is similar to the case of zero sound propagation in single-component neutral Fermi liquid and the magnitude of the slow Fermi surface distortion depends on the relation between $s'$ and $v_1$.
 
%%%%%%%%%%%%%%%%%%%%%%%%%%%%%%%%
\subsection{Acoustic plasmon velocity and damping in three and two dimensions}
\label{Sec:Dispersion}

\begin{figure}[t!]
	\includegraphics[width=\linewidth]{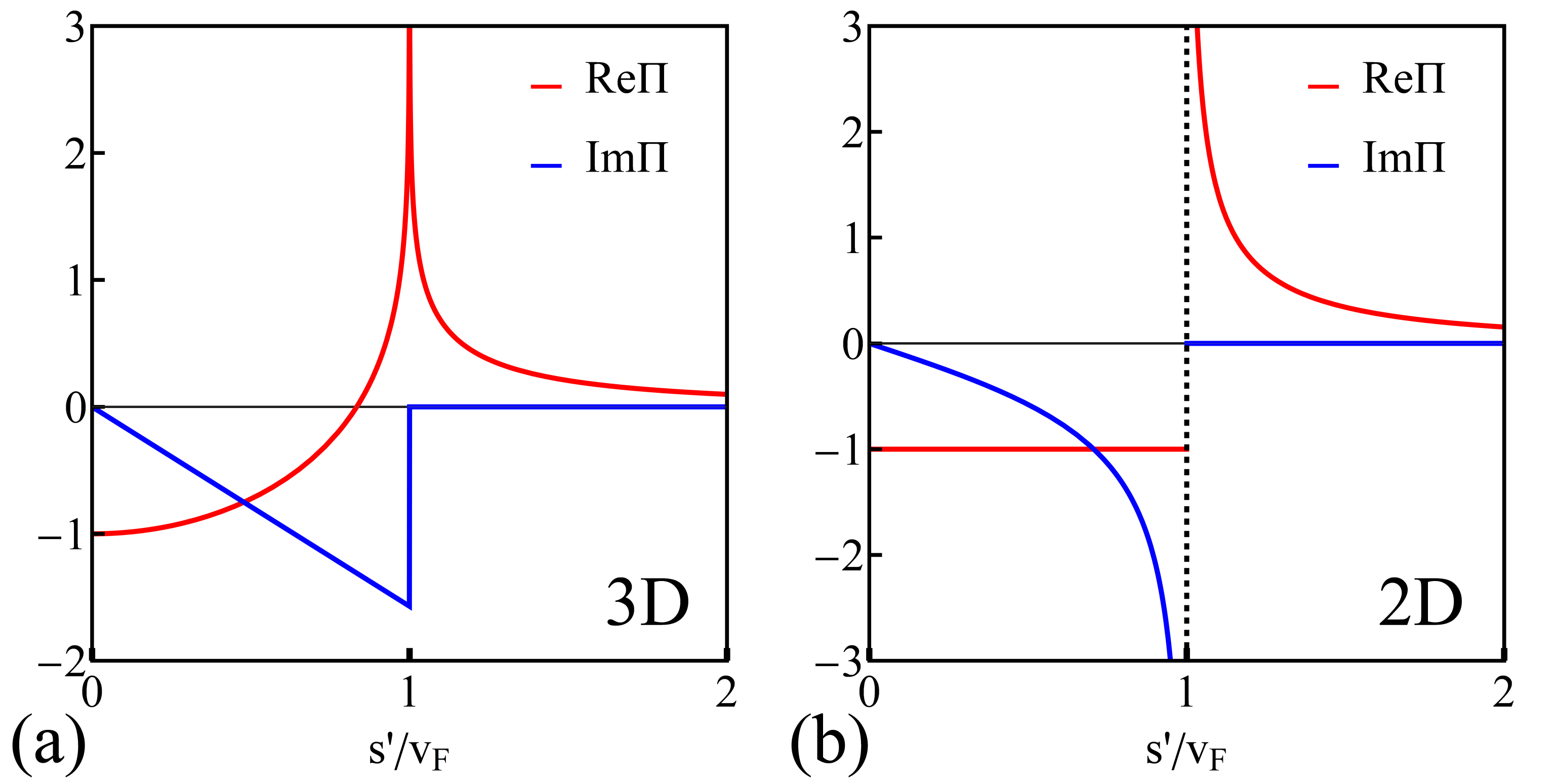}
	\caption{Dimensionless quasi-classical polarizabilities in three and two dimensions after Eqs.~(\ref{Eq:Quasi_P_3D}) and~(\ref{Eq:Quasi_P_2D}).}
	\label{Fig:Quasiclassic_P}
\end{figure}

Now we consider explicit solutions for AP velocity $s'$ and damping $s''$ determined by the specific form of $\tilde{\Pi}_{\alpha}(s)$ for three and two dimensions. In 3D case the quasi-classical polarizability has the form (see Fig.~\ref{Fig:Quasiclassic_P}a)
\begin{equation}
	\label{Eq:Quasi_P_3D}
	\tilde{\Pi}(s)=-1-\frac{s}{2 v_{_F}}\ln\left(\frac{s-v_{_F}}{s+v_{_F}}\right) ,
\end{equation}
and the general equation~(\ref{Eq:AcPl_Disp}) has solution if both $v_2/v_1$ and $F_{\rm eff}$ exceed threshold values. Eq.~(\ref{Eq:AcPl_ZS}) has no general analytic solution in 3D so we focus on its limiting cases. When the effective interaction is weak $F_{\rm eff} \ll 1$, AP velocity coincides with the velocity of zero sound in the almost ideal Fermi gas~\cite{Landau_V9}
\begin{gather}
	\label{Eq:Weak_3D}
	s'=v_1 \left(1+2{\rm e}^{-2/F_{\rm eff}}\right) ,\\
	\frac{s''}{s'}=2\pi\frac{F_{_{\rm RPA}}}{F^2_{\rm eff}}\frac{v_1}{v_2}{\rm e}^{-2/F_{\rm eff}} .
\end{gather}
The propagation of the mode in this regime is accompanied~(see Fig.~\ref{Fig:Shapes}c) by the characteristic sufficient deformation of the slow Fermi surface towards ${\bf k}$~\cite{Landau_V9} since $s'$ is close to $v_1$. In this regime the damping of the mode is small due to both $v_2/v_1 \gg 1$ and $F_{\rm eff}\ll 1$. 
 
In the opposite case of strong interaction $F_{\rm eff} \gg 1$, AP mode is described by Pines plasmon~\cite{Pines1956} which is a solid state analogy of the ion-acoustic wave~\cite{Landau_V10} with
\begin{gather}
	\label{Eq:Strong_3D}
	s'=v_1\sqrt{\frac{F_{\rm eff}}{3}} ,\\
	\frac{s''}{s'}=\frac{\pi}{4}\sqrt{\frac{F_{_{\rm RPA}}}{3 F_{\rm eff}}}\frac{v_1}{v_2} .
\end{gather}
In this regime $v_1 \ll s' \ll v_2$ and the Fermi surface only slightly differs from the equilibrium one as presented in Fig.~\ref{Fig:Shapes}d.

In the case of two dimensions the polarizability has the form (see Fig.~\ref{Fig:Quasiclassic_P}b)
\begin{equation}
	\label{Eq:Quasi_P_2D}
	\tilde{\Pi}(s)=-1+{\rm sign}({\rm Re}\,s)\frac{s}{\sqrt{s^2-v_{_F}^2}} ,
\end{equation}
and the existence of solution of the general equation~(\ref{Eq:AcPl_Disp}) is independent of the ratio $v_2/v_1$~\cite{Vignale1988}. Acoustic plasmon velocity is similar to the zero sound velocity in two dimensions
\begin{gather}
	s'=v_1 \frac{1+ F_{\rm eff}}{\sqrt{1+2 F_{\rm eff}}} ,\\
	\frac{s''}{s'}=\frac{F_{_{\rm RPA}}}{F^2_{\rm eff}}\frac{(s'^2-v_1^2)^{3/2}}{v_1^2\sqrt{v_2^2-s'^2}} .
\end{gather}
In the weak interaction regime $F_{\rm eff}\ll1$
\begin{gather}
	\label{Eq:Weak_2D}
	s'=v_1\left(1+\frac{1}{2}F_{\rm eff}^2\right) ,\\
	\frac{s''}{s'}=F_{_{\rm RPA}} F_{\rm eff} \frac{v_1}{\sqrt{v_2^2-v_1^2}} ,
\end{gather}
and AP is well-defined if the Fermi velocities of components differ strongly. In the strong interaction case $F_{\rm eff}\gg 1$
\begin{gather}
	\label{Eq:Strong_2D}
	s'=v_1\sqrt{\frac{F_{\rm eff}}{2}} ,\\
	\frac{s''}{s'}=\frac{F_{_{\rm RPA}}}{F_{\rm eff}}\frac{s'}{2\sqrt{v_2^2-s'^2}} ,
\end{gather}
and the damping is small when $s'/v_2 \ll 1$.

%%%%%%%%%%%%%%%%%%%%%%%%%
%%%%%%%%%%%%%%%%%%%%%%%%%

Possibility and the particular regime of AP propagation in two-component EL characterized by the specific set of parameters, namely, elementary charges of carriers $e_{1,2}$, mass ratio $m_2/m_1$ and dimensionless Wigner-Seitz radii $r_{1,2}$ depend on the approximation used for the effective interaction parameter $F_{\rm eff}$ or, more generally, for parameters in the dispersion equation~(\ref{Eq:AcPl_Disp}). From now on, indices $1$ and $2$ are used only for numeration and do not indicate the slow and the fast components. Since $v_2/v_1=r_1/r_2$, the theory developed above can be readily applied for $r_1>r_2$ and the index permutation $1\leftrightarrow 2$ is needed when $r_2>r_1$. 

\begin{figure}[t!]
	\includegraphics[width=\linewidth]{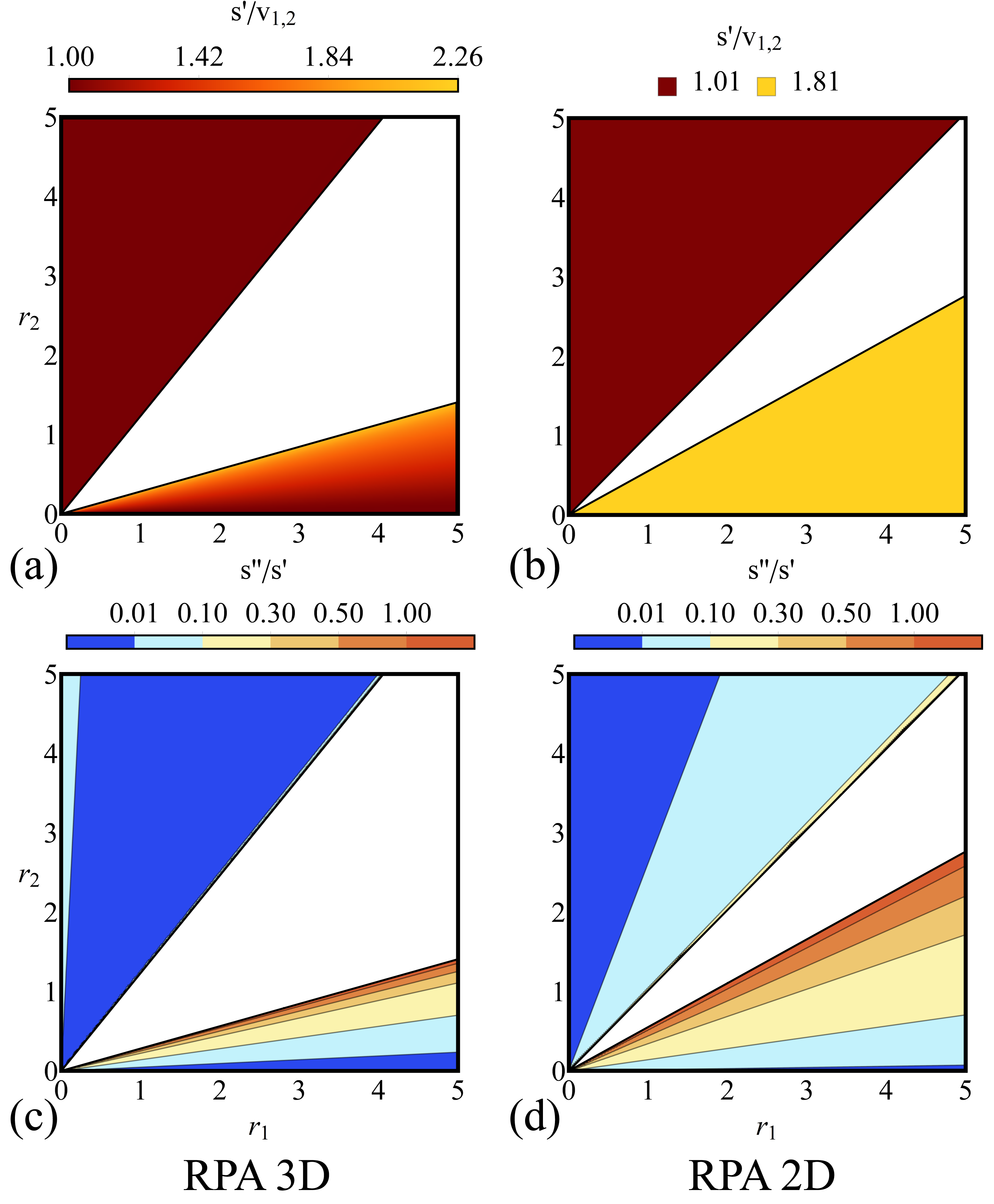}
	\caption{Diagrams of acoustic plasmon velocity (a,b) and damping (c,d) in random phase approximation in three and two dimensions for $m_2/m_1=0.2$.}
	\label{Fig:Diagram_RPA}
\end{figure}

In RPA, propagation of AP is controlled by the ratio of the Thomas-Fermi wave vectors in components $\delta \kappa=\kappa_2/\kappa_1$. Therefore, the case of strong effective interaction is realized when the screening by the fast component is weak and vice versa. Within RPA, APs are equivalent in electron-electron and electron-hole liquids. Diagrams showing the numerically calculated AP velocity and damping according to Eqs.~(\ref{Eq:AcPl_Disp}) and~(\ref{Eq:AP_damping}) at $F_{0}^{\alpha\beta}=0$ and $m_2/m_1=0.2$ are presented in Fig.~\ref{Fig:Diagram_RPA}. In the central parts of the diagrams at $r_2\sim r_1$ there is a forbidden domain where APs do not propagate. The difference in the sizes of the forbidden domains in 2D and 3D is related to different behavior of ${\rm Re}\,\tilde{\Pi}(s)$ for $s'<v_{_F}$ in 3D and 2D (see Fig.~\ref{Fig:Quasiclassic_P}).

In 3D, the RPA effective interaction parameter $F_{_{\rm RPA}}=m_1^2 r_2/m_2^2 r_1$ (for $r_1>r_2$) depends on the ratios of masses and interaction constants. At fixed $m_2/m_1 <1$, the effective interaction is increased (in comparison to the $m_1=m_2$ case) in the lower parts of Figs.~\ref{Fig:Diagram_RPA}a,c and decreases when $r_2>r_1$. In the quasi single-component regions $r_{1,2}/r_{2,1}\ll 1$, $F_{_{\rm RPA}}$ is small due to strong screening by high-density component. Therefore, in the upper domains of the diagrams shown in Fig.~\ref{Fig:Diagram_RPA}a,c acoustic plasmon propagates in the regime of zero sound in the almost ideal Fermi liquid~(\ref{Eq:Weak_3D}). In the lower domains, the effective interaction could be strong near the edge of the forbidden region, far away from the domain of strong screening. However, at the edge of the forbidden domains, AP acquires infinite damping and the mode becomes well-defined with the increase of the Fermi velocities difference. Consequently, the ion-acoustic wave regime~(\ref{Eq:Strong_3D}) is suppressed in 3D two-component ELs even within RPA.

The effective interaction parameter in two dimensions is determined solely by the mass ratio $F_{_{\rm RPA}}=m_1/m_2$ (for $r_1>r_2$). Acoustic plasmon velocity is hence constant in the lower and upper parts of the diagram (Fig.~\ref{Fig:Diagram_RPA}b) and both the well-defined zero sound in the almost ideal Fermi liquid~(\ref{Eq:Weak_2D}) and Pines plasmon~(\ref{Eq:Strong_2D}) regimes take place at $r_2\gg r_1$ and $r_2 \ll r_1$, respectively, provided the mass difference is strong.

%%%%%%%%%%%%%%%%%%%%%%%%%
%%%%%%%%%%%%%%%%%%%%%%%%%

As it is shown above, propagation of AP is favorable when the interaction constants in components are strongly different. Therefore, it is essential to take into account the quasiparticle short-range interaction to study the AP dispersion and damping in two-component ELs. Renormalization of the AP mode and the modification of the RPA diagrams are determined by the behavior of the isotropic Landau parameters in the $r_1-r_2$ plane  
\begin{equation}
	F_0^{\alpha \beta}=F_0^{\alpha \beta}(\delta m, r_1, r_2) ,
\end{equation}
where $\delta m=m_2/m_1$.

%%%%%%%%%%%%%%%%%%%%%%%%%%%%%%%%%%%%%%%%%%%%%%%%%%%%%%%%%%%%%%%%%%%%%%%%%%%%%%%%%%%%%%%%%%%%%%%%%%%%%%%%%%%
%%%%%%%%%%%%%%%%%%%%%%%%%%%%%%%%%%%%%%%%%%%%%%%%%%%%%%%%%%%%%%%%%%%%%%%%%%%%%%%%%%%%%%%%%%%%%%%%%%%%%%%%%%%
\section{Microscopic calculation of isotropic Landau parameters}
\label{Sec:LF}

Short-range quasiparticle interaction in ELs is associated with particle exchange and coulomb correlations. Landau interaction function in two-component EL is defined in the similar manner to the single-component case as a functional derivative of quasiparticle energy with respect to quasiparticle occupation number at Fermi surfaces
\begin{equation}
	\label{Eq:LF_def}
	f^{\alpha\beta}_{\sigma \sigma'}(\vartheta)=\left.\frac{\delta\mathcal{E}_{\alpha}^{\sigma}(\textbf{k})}{\delta n_{\beta}^{\sigma'}(\textbf{k}')}\right|_{\substack{\textbf{k}=\textbf{k}_{\alpha} \\ \textbf{k}'=\textbf{k}_{\beta}'}} ,
\end{equation}
where $\vartheta=\widehat{\textbf{k}_{\alpha} \textbf{k}_{\beta}'}$ is the angle between wave vectors of quasiparticles at Fermi surfaces,  $\sigma$ and $\sigma'$ are spin indicies. Microscopic calculation of $f_{\sigma \sigma'}^{\alpha\beta}(\vartheta)$ is based on~(\ref{Eq:LF_def}) and the particular form of the interaction-induced renormalization of single particle spectrum within quantum field theory~\cite{Rice1965,Hedin1965,Ting1975,Yarlagadda1994,Giuliani_Vignale_2005}.

In this work we use on-shell approximation for quasiparticle energy given by
\begin{equation}
	\label{Eq:OSA}
	\mathcal{E}_{\alpha}^{\sigma}=\epsilon_{\alpha}^{\sigma}(\textbf{k})+\mathrm{Re}\,\Sigma_{\alpha}^{\sigma}(\textbf{k},\epsilon_{\alpha}^{\sigma}(\textbf{k})) ,
\end{equation}
where $\epsilon_{\alpha}^{\sigma}(\textbf{k})$ is the energy dispersion of non-interacting fermions of type $\alpha$. The self energy $\Sigma_{\alpha}^{\sigma}$ is calculated in the $G_0W$-approximation 
\begin{equation}
	\label{Eq:SelfEnergy}
	\Sigma_{\alpha}^{\sigma}(\textbf{k},\omega)=i\sum_{\textbf{q}\omega'}W(\textbf{q},\omega')G_{\alpha}^{\sigma}(\textbf{k}-\textbf{q},\omega-\omega') ,
\end{equation}
where $\sum_{\textbf{q}\omega'}=(2\pi)^{-d-1}\int d\textbf{q}d\omega'$, $G_{\alpha}^{\sigma}(\textbf{k},\omega)=\left(\omega-\epsilon_{\alpha}^{\sigma}(\textbf{k})+i\eta_{\textbf{k}}\right)^{-1}$ is the Green function of non-interacting fermions and $\eta_{\textbf{k}}=\eta\,\mathrm{sgn}(k-k_{\alpha})$ is infinitesimal ($\eta\to 0_{+}$). For the effective interaction $W(\textbf{q},\omega)$ the random-phase approximation (RPA) is used
\begin{equation}
	\label{Eq:Weff}
	W(\textbf{q},\omega)=\frac{V(\textbf{q})}{\varepsilon_{_{\mathrm{RPA}}}(\textbf{q},\omega)} ,
\end{equation}
where $V(\textbf{k})$ is the Fourier transform of the bare Coulomb interaction and RPA dielectric function has the form 
\begin{gather}
	\varepsilon_{_{\mathrm{RPA}}}(\textbf{q},\omega)=1-\sum\limits_{\alpha \sigma} V(\textbf{q})\Pi_{\alpha}^{\sigma}(\textbf{q},\omega) ,\\
	\label{Eq:RPA_P}
	\Pi_{\alpha}^{\sigma}(\textbf{q},\omega)=-i\sum\limits_{\textbf{k}'\omega'}G_{\alpha}^{\sigma}(\textbf{k}',\omega')G_{\alpha}^{\sigma}(\textbf{q}+\textbf{k}',\omega+\omega') .
\end{gather}
Here $\Pi_{\alpha}^{\sigma}(\textbf{k},\omega)$ is the non-interacting polarizability of the fermions of type $\alpha$.

We consider isotropic, non-polarized system with two types of fermions residing in the vicinity of extremums of two weakly overlapping parabolic bands. This is consistent with the band structure of transition metals with incomplete inner shell~\cite{Frohlich1968}, photoexcited electron-hole liquids in pure semiconductors~\cite{Giuliani_Vignale_2005} and elemental bismuth~\cite{Ruhman2017}. Due to different symmetries of the band wave functions, overlap integrals are diagonal in component indices $|\langle\alpha\sigma \textbf{k}|\beta\sigma' \textbf{k}'\rangle|^2\approx \delta_{\alpha\beta}\delta_{\sigma\sigma'}$, so the contribution of intercomponent transitions to the self energy~(\ref{Eq:SelfEnergy}) and RPA polarizability~(\ref{Eq:RPA_P}) is negligible. In the case of bismuth they are also accompanied by the transfer of large wave vector comparable to the size of the Brillouin zone.

Using Eqs.~(\ref{Eq:LF_def}), (\ref{Eq:OSA}) and (\ref{Eq:SelfEnergy}) and following the method described in~\cite{Rice1965,Ting1975,Yarlagadda1994}, we obtain the spin-symmetric part of the Landau interaction function $f^{\alpha\beta}_s=(f^{\alpha\beta}_{\uparrow\uparrow}+f^{\alpha\beta}_{\uparrow\downarrow})/2$ given by two terms
\begin{equation}
	\label{Eq:LF_tot}
	f^{\alpha\beta}_s(\vartheta)=f^{\alpha\beta}_{\mathrm{SX}}(\vartheta)+f^{\alpha\beta}_{\mathrm{CH}}(\vartheta) .
\end{equation}
The first term is the screened exchange (SX) contribution
\begin{equation}
	\label{Eq:LFsx}
	f^{\alpha\beta}_{\mathrm{SX}}(\vartheta)=-\frac{1}{2}\delta_{\alpha\beta}W(\textbf{k}_{\alpha}-\textbf{k}_{\beta}',0) ,
\end{equation}
determined by Pauli repulsion of non-interacting indistinguishable fermions with the same spin projection (exchange hole). Consequently, it is diagonal in components indices and proportional to the screened interaction at the Fermi surface. The second term is the coulomb hole (CH) contribution determined by particle correlations due to coulomb interaction
\begin{multline}
	\label{Eq:LFch}
	f^{\alpha\beta}_{\mathrm{CH}}(\vartheta)=-\mathrm{Re}\,\sum\limits_{\textbf{q}\omega}W^2(\textbf{q},i\omega)G_{\alpha}(\textbf{k}_{\alpha}-\textbf{q},\mu_{\alpha}-i\omega)\times \\
	\times\left[G_{\beta}(\textbf{k}_{\beta}'+\textbf{q},\mu_{\beta}+i\omega)+G_{\beta}(\textbf{k}_{\beta}'-\textbf{q},\mu_{\beta}-i\omega)\right] .
\end{multline}
It is second order by $W(\textbf{k},\omega)$ and both the scattering processes at the Fermi surfaces and outside them contribute to $f_{\mathrm{CH}}^{\alpha\beta}(\vartheta)$. Intercomponent quasiparticle interaction is determined only by the CH term. 

Equations~(\ref{Eq:LFsx}) and~(\ref{Eq:LFch}) show that short-range exchange-correlation interaction in two-component ELs is attractive (similarly to the single-component case) and, in contrast to the long-range coulomb interaction, is independent of the signs of particles' charge. This asymmetry can be understood by qualitative comparison of correlation holes in electron-electron and electron-hole liquids. Electron-electron case is similar to the single-component one. Electron under scrutiny is surrounded by the depletion area of positive spatial charge due to intra- and intercomponent coulomb repulsion. The resulting short-range interaction with this correlation hole is attractive. In electron-hole liquids, vicinity of the electron under scrutiny is depleted with the other electrons and is populated by holes. Again, the correlation hole has positive effective charge and short-range interaction with it is attractive.

To obtain isotropic Landau parameters, interaction function~(\ref{Eq:LF_tot}) should be additionally averaged over the directions of $\textbf{k}'$: $F_0^{\alpha\beta}=D_{\beta}\sum_{\Omega_{\textbf{k}'}}f_s^{\alpha\beta}(\vartheta)$. Note that in~(\ref{Eq:LFch}) we have transformed the $\omega$-integration to imaginary axis~\cite{Rice1965,Hedin1965,Ting1975,Yarlagadda1994,Giuliani_Vignale_2005} to avoid plasmon poles and the poles of $G_{\alpha}(\textbf{k},\omega)$. RPA dielectric function is real-valued at imaginary frequencies and is not associated with any particle-hole and collective density excitations. Consequently, the only characteristic scale which determines the screening strength by the $\alpha$ component in~(\ref{Eq:Weff}) is the Thomas-Fermi wave vector $\kappa_{\alpha}$. For isotropic system, angular integrations appearing in the CH term can be carried out analytically and the expressions for the SX and the CH contributions to $F_0^{\alpha\beta}$ take the forms~(\ref{Apdx:F3sx}),~(\ref{Apdx:F3ch}) in 3D and~(\ref{Apdx:F2sx}),~(\ref{Apdx:F2ch}) in 2D suitable for numerical calculations. The intra- and intercomponent Landau parameters are connected via a simple permutation of the arguments 
\begin{gather}
	F_{0}^{22}(\delta m, r_1, r_2)=F_{0}^{11}(\delta m^{-1}, r_2, r_1) ,\\
	F_{0}^{12}(\delta m, r_1, r_2)=F_{0}^{21}(\delta m^{-1}, r_2, r_1) .
\end{gather}
Consequently, further we will focus only on $F_0^{11}$ and $F_0^{21}$. 

%%%%%%%%%%%%%%%%%%%%%%%%%
\begin{figure*}[t!]
	\includegraphics[]{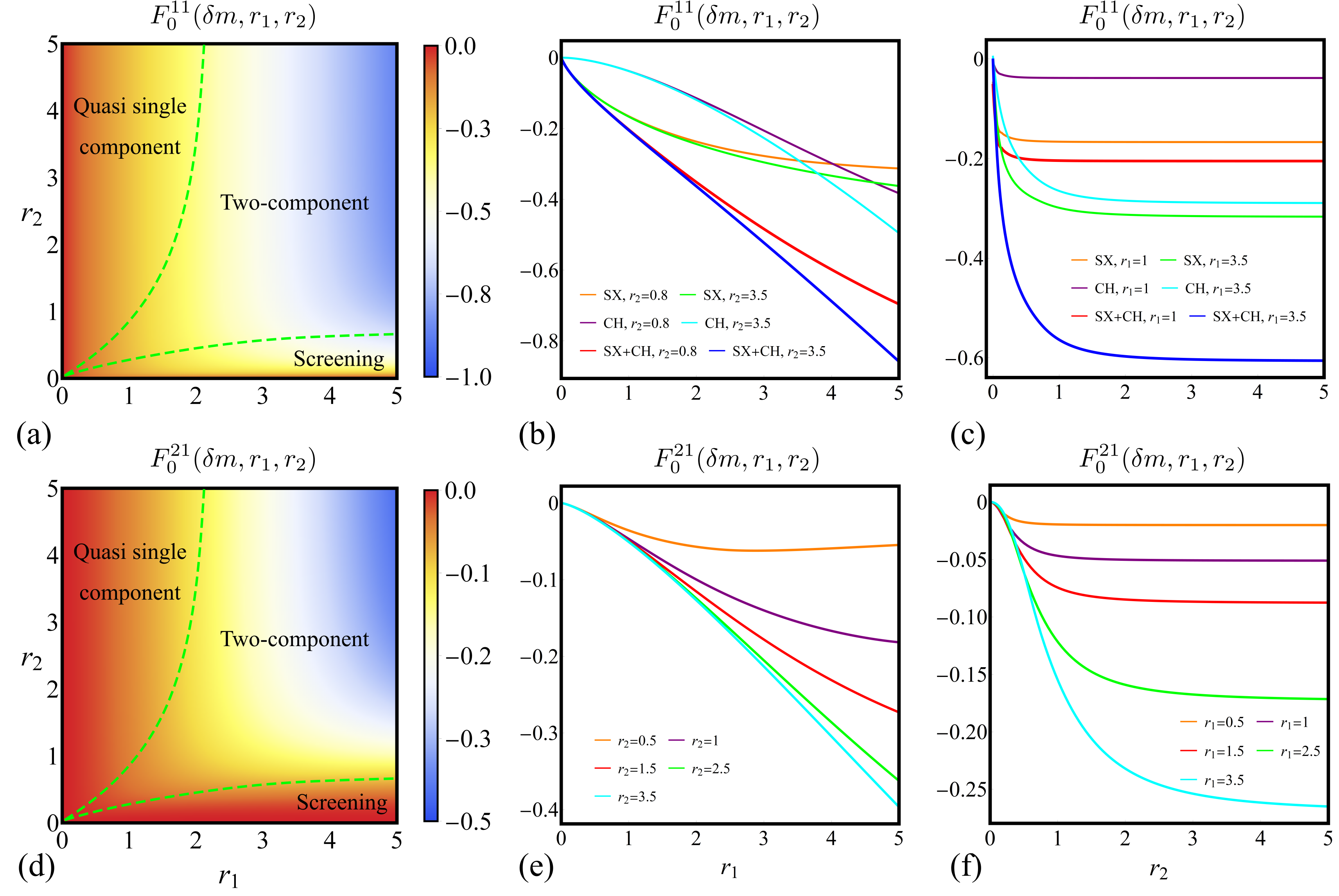}
	\caption{Isotropic Landau parameters in three-dimensional two-component electron liquid at $m_2/m_1=0.2$. Panels (a) and (d) show the density plots of intra- and intercomponent parameters $F_0^{11}$ and $F_0^{21}$. Their cuts and partial screened exchange (SX) and coulomb hole (CH) contributions at fixed $r_1$ and $r_2$ are presented on panels (b), (c), (e) and (f), respectively.}
	\label{Fig:LF3}
\end{figure*}
%%%%%%%%%%%%%%%%%%%%%%%%%

At fixed mass ratio, $r_1-r_2$ plane is divided into several characteristic domains of quasiparticle interaction. In 3D, the ratio of Thomas-Fermi wave vectors $\kappa_2/\kappa_1=\delta m/\sqrt{\delta r_s}$ depends on the ratios of masses $\delta m=m_2/m_1$ and interaction constants $\delta r_s=r_2/r_1$. Therefore, the relative screening strength of the coulomb interaction at the first Fermi surface induced by components varies with $r_1$ and $r_2$. At high densities $r_{1,2}\ll 1$, isotropic Landau parameters take the following analytic form when $\kappa_2/\kappa_1\lesssim 1$
\begin{gather}
	\label{Eq:F3sx_hd}
	F_{0,\mathrm{SX}}^{11}=\frac{\gamma_3 r_{1}}{2\pi}\ln\left(\frac{\gamma_3 r_{1}}{\pi}\frac{\kappa_1^2+\kappa_2^2}{\kappa_{1}^2}\right) ,\\
	\label{Eq:F3ch_hd}
	F_{0,\mathrm{CH}}^{\alpha 1}=\frac{\gamma_3^2 r_{\alpha}^2}{\pi^2} \frac{m_{1}}{m_{\alpha}} \mathcal{I}_3\left(\frac{r_{\alpha}}{r_{1}}\right) \ln\left(\frac{\gamma_3 r_{1}}{\pi}\frac{\kappa_1^2+\kappa_2^2}{\kappa_{1}^2}\right) ,
\end{gather}
where $\gamma_3=(4/9\pi)^{1/3}$ and $\mathcal{I}_3(x)=1/(1+x)$. At low $r_{1,2}$, the screening length is greater than interparticle distance in components $\lambda_{TF} \gg \lambda_{F}^{\alpha}$ and Landau parameters are determined by interaction at the Fermi surfaces $F_{c}^{\alpha\alpha}(k_{\alpha})\sim r_{\alpha}$ and by the coulomb logarithm associated with the screened processes of forward scattering. In a wider range of particle densities in components, Landau parameters can be calculated numerically using Eqs.~(\ref{Apdx:F3sx}) and~(\ref{Apdx:F3ch}). The results for $m_2/m_1=0.2$ are demonstrated in Fig.~\ref{Fig:LF3}. It can be clearly seen that the intercomponent interaction is comparable to the intracomponent one. 

On the left sides of Figs.~\ref{Fig:LF3}a and~\ref{Fig:LF3}d, there is a region where $\mathrm{\max}[1,r_1]\, \kappa_2^2/\kappa_1^2 \ll 1$ (i.e. $r_2\gg \mathrm{max}[r_1,r_1^2]\, \delta m^2$) and the effective interaction in~(\ref{Apdx:F3sx}) and~(\ref{Apdx:F3ch}) is determined by screening by the high-density first component. As it follows from~(\ref{Eq:F3sx_hd}),~(\ref{Eq:F3ch_hd}) and Figs.~\ref{Fig:LF3}b and~\ref{Fig:LF3}c, intracomponent part of the short-range interaction $F_0^{11}$ in this domain replicates the case of the isolated first component being almost independent of $r_2$. This quasi single-component region is characterized by low $r_1$ and, consequently, weak quasiparticle interaction $F_{0}^{11},F_{0}^{21}\ll 1$ vanishing in the $r_1\to 0$ limit.

Two-component domain is characterized by comparable Thomas-Fermi wave vectors $\kappa_2 \sim \kappa_1$. In this region, RPA dielectric function is determined by both components and the variation of Landau parameters with $r_{1,2}$ can be traced in terms of averaged interaction parameter $\rho_s=\sqrt{r_1^2+r_2^2}$. In high density limit $\rho_s \ll 1$, Landau parameters are described by Eqs.~(\ref{Eq:F3sx_hd}),~(\ref{Eq:F3ch_hd}) and the quasiparticle interaction is weak. With the decease of particle density $\rho_s \geq 1$, screening length becomes the shortest one $\lambda_{TF} \ll \lambda_{F}^{\alpha}$ and the effective particle interaction at the Fermi surfaces is significantly reduced. Screened exchange term saturates at these densities while the absolute value of the CH one grows with $\rho_s$ as it is determined by scattering processes outside the Fermi surfaces and dynamical screening, which is much weaker than the static one (see Appendix).

On the bottom of Figs.~\ref{Fig:LF3}a and~\ref{Fig:LF3}d, there is a screening domain at $r_2\ll \mathrm{min}[r_1,r_1^2]\, \delta m^2$. Intra- and intercomponent short-range interaction are weak (see Figs.~\ref{Fig:LF3}c and~\ref{Fig:LF3}f) due to strong screening by the high-density second component, which determines the RPA dielectric function. At $r_1 \ll 1$, isotropic Landau parameters in the screening domain take the form 
\begin{gather}
	F_{0,\mathrm{SX}}^{11}=-\frac{r_2}{2 r_1}\frac{m_1^2}{m_2^2} ,\\
	F_{0,\mathrm{CH}}^{11}=-0.6 \sqrt{\frac{\gamma_3}{\pi}}\frac{r_2^{3/2}}{\pi r_1}\frac{m_1^3}{m_2^3} ,\\
	F_{0,\mathrm{CH}}^{21}=-1.2 \left(\frac{\gamma_3}{\pi}\right)^{3/2}\frac{r_2^{5/2}}{\pi r_1}\frac{m_1^2}{m_2^2} .
\end{gather}
Both $F_0^{11}$ and $F_0^{21}$ vanish in the limit $r_2\to 0$.

%%%%%%%%%%%%%%%%%%%%%%%%%
\begin{figure*}[t!]
	\includegraphics[]{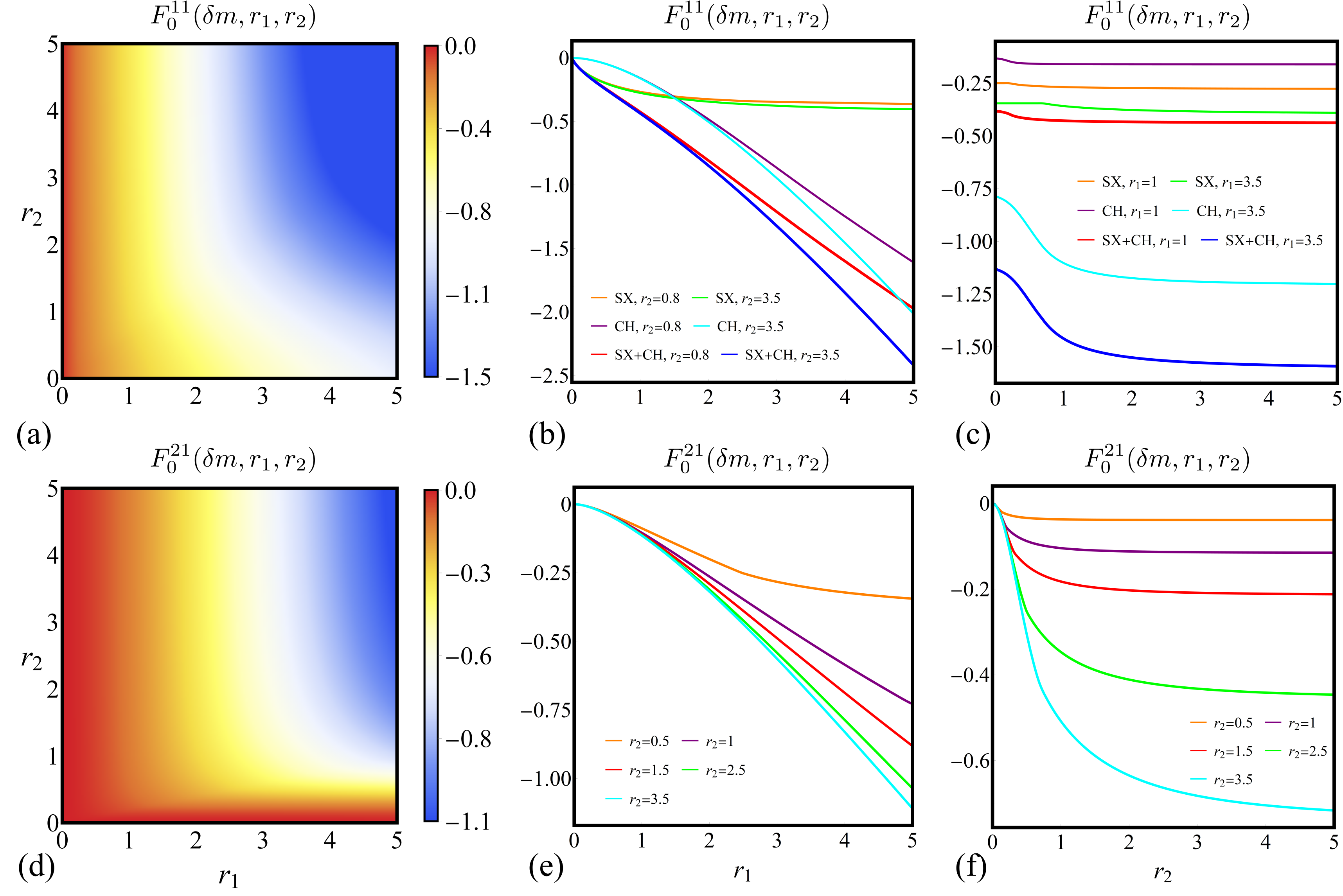}
	\caption{Isotropic Landau parameters in two-dimensional two-component electron liquid at $m_2/m_1=0.2$. Panels (a) and (d) show the density plots of intra- and intercomponent parameters $F_0^{11}$ and $F_0^{21}$. Their cuts and partial screened exchange (SX) and coulomb hole (CH) contributions at fixed $r_1$ and $r_2$ are presented on panels (b), (c), (e) and (f), respectively.}
	\label{Fig:LF2}
\end{figure*}
%%%%%%%%%%%%%%%%%%%%%%%%%

In two-dimensional case, the ratio of Thomas-Fermi wave vectors $\kappa_2/\kappa_1=m_2/m_1$ is determined by the mass difference only and the relative screening strength by components is constant in the whole $r_1-r_2$ plane. In the high density limit $r_{1,2}\ll 1$, Landau parameters take the form
\begin{gather}
	\label{Eq:F2sx_hd}
	F_{0,\mathrm{SX}}^{\alpha \beta}=\delta_{\alpha \beta}\frac{\gamma_2 r_{\alpha}}{\pi}\ln\left(\gamma_2 r_{\alpha} \frac{\kappa_1+\kappa_2}{\kappa_{\alpha}}\right) ,\\
	\label{Eq:F2ch_hd}
	F_{0,\mathrm{CH} }^{\alpha \beta}=-\frac{\gamma_2^2 r_{\alpha}^2}{\pi^2} \frac{m_{\beta}}{m_{\alpha}} \mathcal{I}_2\left(\frac{r_{\alpha}}{r_{\beta}}\right) ,
\end{gather}
where $\mathcal{I}_2(x)=\frac{2x}{(1-x^2)^2}\left[(1+x^2)K(1-x^2)-2 E(1-x^2)\right]$, $K(x)$ and $E(x)$ are the complete elliptic integrals of the first and second kinds~\cite{Elliptic_Integrals_Note}, respectively, and  $\gamma_2=1/\sqrt{2}$. In contrast to 3D case, CH contribution in 2D is not affected by the coulomb logarithm. According to Eqs.~(\ref{Eq:F2sx_hd}) and~(\ref{Eq:F2ch_hd}), in two-dimensional ELs intracomponent interaction qualitatively reproduce the single- component case with constant renormalization controlled by the mass ratio. This is consistent with the results of numerical calculation of Landau parameters presented in Fig.~\ref{Fig:LF2}. As it is shown in Figs.~\ref{Fig:LF2}a-c, the dependence of $F_0^{11}$ on $r_2$ is very weak and $F_0^{11}$ is nonzero in the $r_2\to 0$ limit. The intercomponent Landau parameter $F_0^{21}$ vanishes in the single-component limits $r_{1,2}\to 0$ and its values are comparable to $F_0^{11}$ at $r_{1,2}\geq 1$, see Figs.~\ref{Fig:LF2}d-f. Similarly to the single-component case, short-range interaction is stronger in 2D than in 3D.

The remaining parameters $F_0^{22}$ and $F_0^{12}$ associated with the second component are of the same order as $F_{0}^{11}$ and $F_{0}^{21}$ when $\delta m \approx 1$ and much smaller when $\delta m\ll 1$.

%%%%%%%%%%%%%%%%%%%%%%%%%%%%%%%%%%%%%%%%%%%%%%%%%%%%%%%%%%%%%%%%%%%%%%%%%%%%%%%%%%%%%%%%%%%%%%%%%%%%%%%%%%%
%%%%%%%%%%%%%%%%%%%%%%%%%%%%%%%%%%%%%%%%%%%%%%%%%%%%%%%%%%%%%%%%%%%%%%%%%%%%%%%%%%%%%%%%%%%%%%%%%%%%%%%%%%%
\section{Renormalized acoustic plasmons}
\label{Sec:Renorm_AP}

%%%%%%%%%%%%%%%%%%%%%%%%%
\begin{figure*}[t!]
	\includegraphics[width=1.0\linewidth]{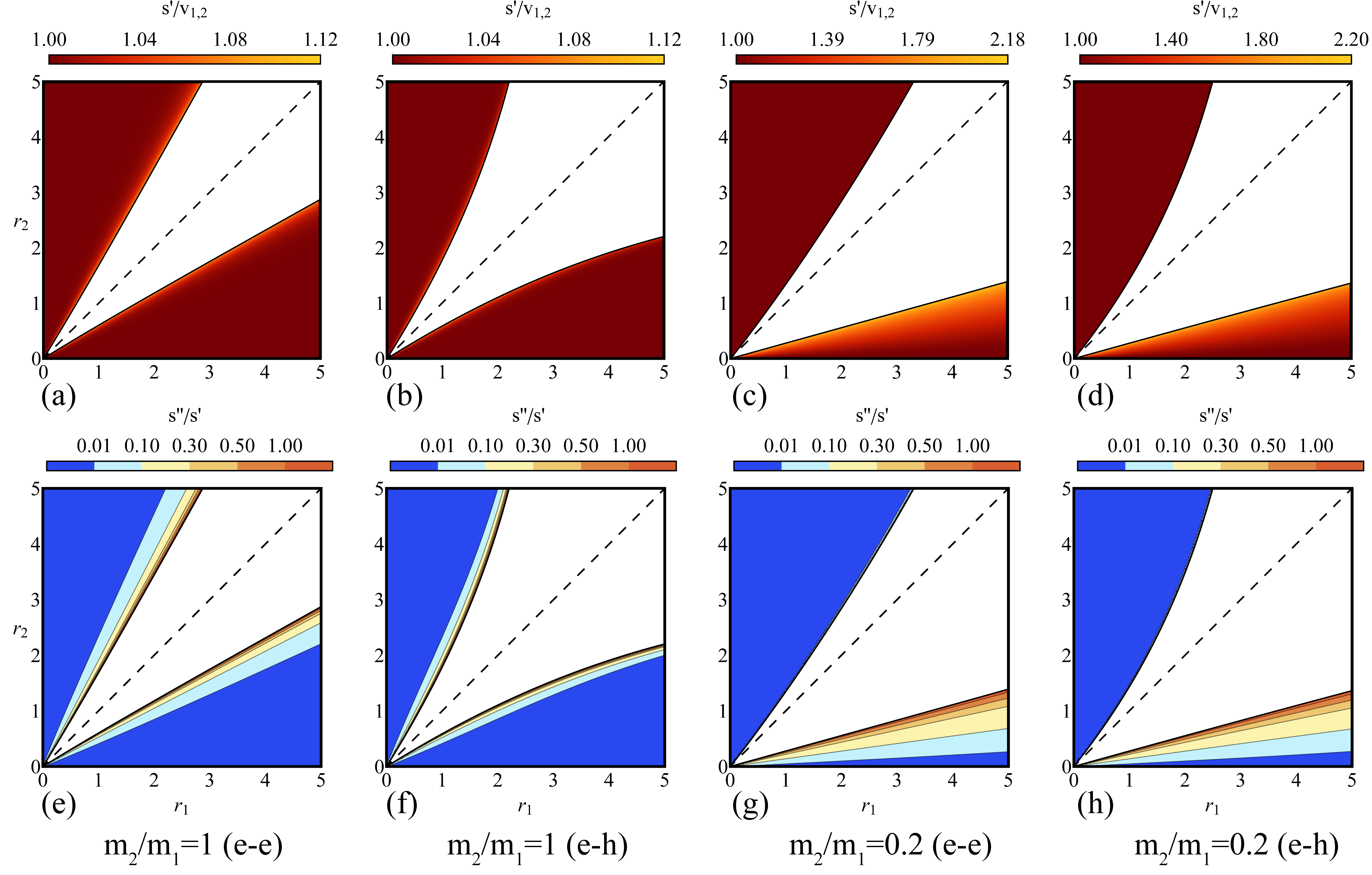}
	\caption{Diagrams of acoustic plasmon velocity (a-d) and damping (e-h) in three-dimensional two-component electron-electron and electron-hole liquids with equal and strongly different masses in components.}
	\label{Fig:AP_renorm_3D}
\end{figure*}
%%%%%%%%%%%%%%%%%%%%%%%%%

\begin{figure*}[t!]
	\includegraphics[width=1.0\linewidth]{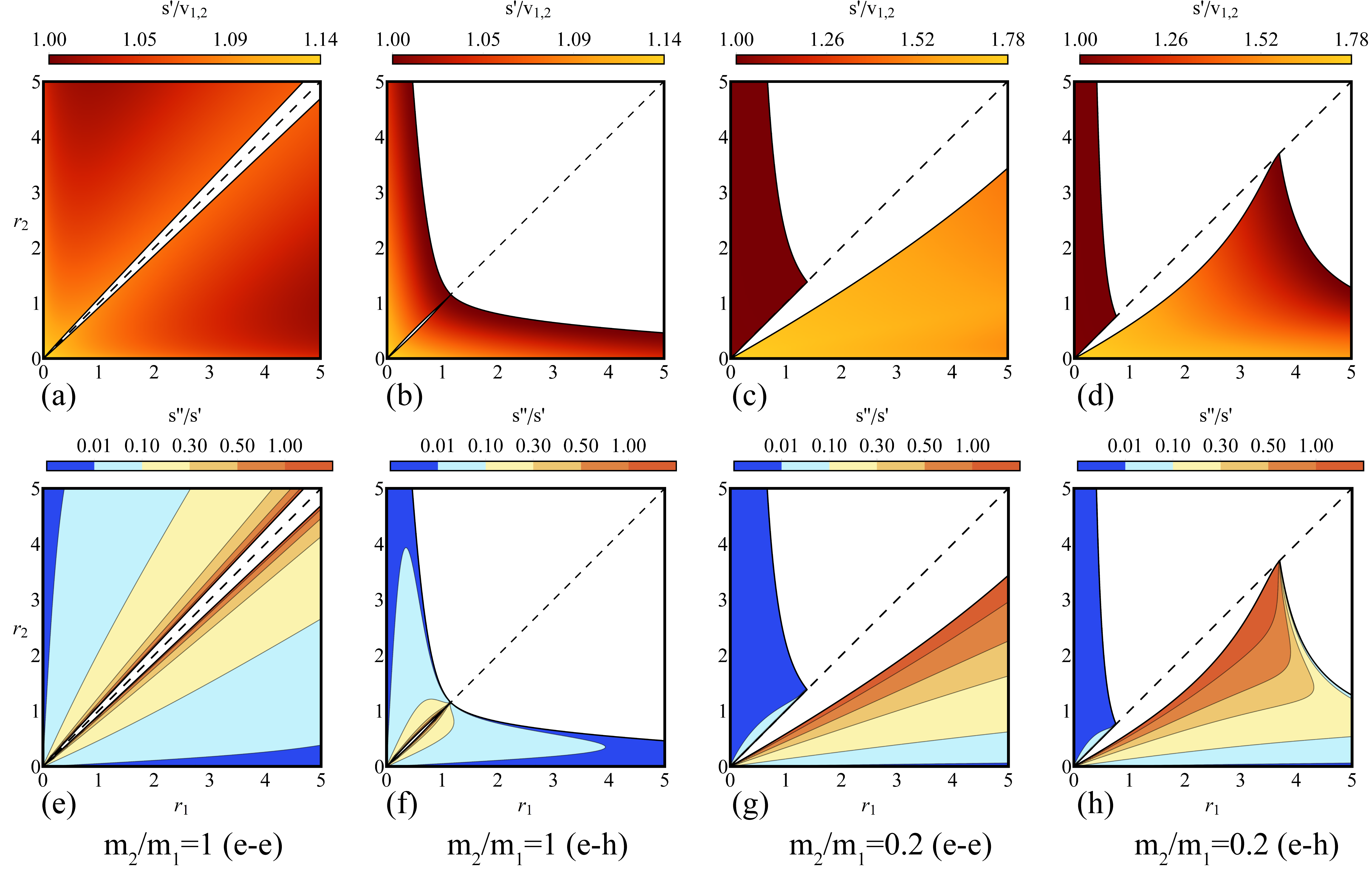}
	\caption{Diagrams of acoustic plasmon velocity (a-d) and damping (e-h) in two-dimensional two-component electron-electron and electron-hole liquids with equal and strongly different masses in components.}
	\label{Fig:AP_renorm_2D}
\end{figure*}

Now we can use the numerically calculated Landau parameters from the previous section to determine the AP renormalization in two-component ELs. The diagrams of AP velocity and damping given by Eq.~(\ref{Eq:AcPl_Disp}) for the cases of electron-electron and electron-hole liquids with equal and strongly different masses in components are presented in Fig.~\ref{Fig:AP_renorm_3D} and Fig.~\ref{Fig:AP_renorm_2D} for 3D and 2D, respectively.

It is seen from Figs.~\ref{Fig:AP_renorm_3D} and~\ref{Fig:AP_renorm_2D} that APs in electron-electron and electron-hole liquids are non-equivalent [see the expression for the reduced Landau parameters~(\ref{Eq:Reduced_F}) entering the dispersion equation~(\ref{Eq:AcPl_Disp})]. This is a direct consequence of the asymmetry of intercomponent short-range interaction with respect to the sign of elementary charges of particles in components. This peculiarity can be qualitatively understood by considering the restoring forces $\bm{\mathcal{F}}_{\alpha}(\textbf{r},\textbf{p},t)=-\nabla_{\textbf{r}}\mathcal{E}_{\alpha}(\textbf{r},\textbf{p},t)$ which act on the quasiparticles of type $\alpha$ in the vibration. The complex amplitude of the partial contribution to $\bm{\mathcal{F}}_{\alpha}$ exerted due to density variation $\delta N_{s}^{(\beta)}$ in the $\beta$ component is
\begin{equation}
	\label{Eq:Force_partial}
	\bm{\mathcal{F}}_{\omega\textbf{k}}^{\alpha\beta}=-i\textbf{k}\left[V_{\alpha\beta}(k)+f_0^{\alpha\beta}\right]\delta N_{s}^{(\beta)} ,
\end{equation}
where $f_0^{\alpha\beta}$ is the isotropic part of the Landau interaction function~(\ref{Eq:LF_tot}). In the long-wavelength domain the partial restoring force $\bm{\mathcal{F}}_{\omega\textbf{k}}^{\alpha\beta}$ is dominated by the long-range electrostatic interaction and the phase relations between the quasiparticle density oscillations in components~(\ref{Eq:Phase_relation}) [see also Fig.~\ref{Fig:Scheme}a,b] guarantee its cancellation in the total force
\begin{equation}
	\label{Eq:Force_total}
	\bm{\mathcal{F}}_{\omega\textbf{k}}^{\alpha}=-i\textbf{k}\left(f_0^{\alpha\alpha}-\frac{e_{\alpha}}{e_{\beta}}f_0^{\alpha\beta}\right)\delta N_{s}^{(\alpha)} ,\,\alpha\neq\beta.
\end{equation} 
Note that the total restoring forces in the AP mode are determined by the short-range quaisparticle interaction just like the case of conventional zero sound in single-component neutral Fermi liquids. According to Eqs.~(\ref{Eq:Force_partial}) and~(\ref{Eq:Force_total}), intracomponent short-range attraction counteracts the long-range electrostatic repulsion while the intercomponent short-range interaction is responsible for the established charge asymmetry.

Attractive quasiparticle interaction reduces the effective interaction parameter~(\ref{Eq:Feff}) in comparison to the RPA case $F_{\rm eff}< F_{_{\rm RPA}}$. Consequently, one could expect the reduction of the AP velocity $s'<s'_{_{\rm RPA}}$ and the increase of damping $s''>s''_{_{\rm RPA}}$ in two-component ELs. This is consistent with the results of calculation performed in~\cite{Vignale1985,Vignale1988} for electron-hole liquids within the generalized RPA. This approximation is equivalent to using the Eq.~(\ref{Eq:AcPl_Disp}) for the AP dispersion with the reduced Landau parameters of the form $\tilde{F}^{\alpha\alpha}_0=-\frac{2}{\pi}\gamma_d r_{\alpha}$ when the static local field factors are chosen to satisfy the compressibility sum rule. 

Behavior of the microscopically calculated isotropic Landau parameters is more complicated and the resulting AP renormalization differs from~\cite{Vignale1985,Vignale1988}. In 3D the main effect of the short-range interaction is the reconstruction of the forbidden domain at $r_1 \sim r_2$, where the strength of the short-range interaction grows with $\rho_s=\sqrt{r_1^2+r_2^2}$. In the quasi single-component domains both $F_{_{\rm RPA}}$ and $F_0^{\alpha\beta}$ are small due to strong screening by high-density component and the effect of renormalization is minimal: ion-acoustic wave regime is still suppressed (like in RPA) and the well-defined AP propagates in the weakly interacting regime at $r_2\ll r_1$ and $r_2 \gg r_1$ independently on the mass ratio, see Fig.~\ref{Fig:AP_renorm_3D}.

%%%%%%%%%%%%%%%%%%%%%%%%%
%%%%%%%%%%%%%%%%%%%%%%%%%

In two dimensions, the effects of charge asymmetry and the reconstruction of the forbidden domain are more prominent~(see Fig.~\ref{Fig:AP_renorm_2D}) than in 3D since $F_{_{\rm RPA}}$ is constant. The variation of AP velocity with $r_{1,2}$ is determined by the behavior of the isotropic Landau parameters in the $r_1-r_2$ plane. The renormalization of the AP mode is minimal, like in 3D, since the domains where the mode is well-defined coincide with the regions of strong screening. Pines plasmon can propagate in 2D two-component ELs with the strong mass difference when the heavier component is the slowest and the weakly interacting propagation regime is realized when it is the fastest.

The values of the isotropic Landau parameters used to establish the renormalization of the AP mode were calculated in Section~\ref{Sec:LF} in the random-phase approximation, which is exact only in the high density domains $r_{1,2}\ll 1$. The higher quantitative accuracy can be achieved by the quantum Monte Carlo (QMC) or generalized RPA calculations~\cite{Giuliani_Vignale_2005}. However, the former are rather complicated while the accuracy of the latter strongly depends on the approximations used for the local field factors~\cite{Giuliani_Vignale_2005} and their generalization for the two-component case is the subject of the separate work. However, we stress that the calculated diagrams in 3D and 2D describe the actual picture qualitatively well and the main results of this section, namely, charge asymmetry and the stability of the AP mode to the short-range interaction are valid regardless of the approximation used. In addition to this, an excellent convergence of the RPA and QMC results for the compressibility of 3D EL~\cite{Simion2008} determined by $F_0$ indicates that our calculation in 3D can be accurate beyond the high-density domain.

%%%%%%%%%%%%%%%%%%%%%%%%%%%%%%%%%%%%%%%%%%%%%%%%%%%%%%%%%%%%%%%%%%%%%%%%%%%%%%%%%%%%%%%%%%%%%%%%%%%%%%%%%%%
%%%%%%%%%%%%%%%%%%%%%%%%%%%%%%%%%%%%%%%%%%%%%%%%%%%%%%%%%%%%%%%%%%%%%%%%%%%%%%%%%%%%%%%%%%%%%%%%%%%%%%%%%%%
\section{Discussion and conclusions}
\label{Sec:Discussion}

The main reason why APs in two-component ELs have been attracting significant interest over decades is their ability to mediate the electron-electron interaction in the fast component. The resulting effective interparticle interaction of the Fr\"ohlich type opens up possibilities for the formation of plasmonic superconductivity and the emergence of the acoustic plasmaron~\cite{Lundqvist1967,Principi2011} features in the single-particle spectrum of the fast component. However, the reliable experimental manifestation of these phenomena is still missing. The strength of the effective electron-electron interaction depends on the electron-acoustic plasmon coupling constant $|g_k|^2\sim (s'^2/v_1^2-1)^{3/d}$. Therefore, the abovementioned effects are favorable only if AP propagates in the ion-acoustic wave regime, which is similar to the renormalized acoustic phonon in the jellium model~\cite{Mattuk_1992}. However, as we have seen above, the Pines plasmon regime is suppressed in 3D and thus the associated Fr\"ohlich-like interaction is missing. In two dimensions, the AP mediated interaction is weak since the corresponding threshold for the effective interaction constant $F_{\rm eff}\gg 8$ needs $m_2/m_1 \ll 0.1$ which is hardly achievable in the two-component liquids with parabolic dispersion.

Beyond long wavelength domain considered in this work, dispersion of AP becomes non-linear and converges to the boundary of the particle-hole continuum of the slow component. The $k$-dependent velocity of AP at shorter wavelengths $s'(k)< s'_0=\lim\limits_{k\to 0} s'(k)$ is lower then its value in the long wavelength limit $s'_0$. Consequently, electron-acoustic plasmon coupling strength in the fast component is also weaker. Since the calculated velocities $s'_0(\delta m,r_1,r_2)$ represent the upper estimates for $s'(k)$, the results of the paper including suppression of the ion-acoustic wave regime and the associated weakness of the effective electron-electron interaction hold for the entire quasi-classical domain $k\ll k_{1,2}$ where the kinetic equations~(\ref{Eq:KinEq}) for quasiparticles are valid. 

The most promising platform where both regimes of AP propagation manifest and the mediated electron-electron interaction can be strong enough are anisotropic multivalley semimetals. In these materials, quasiparticles which reside in the vicinity of nonequivalent valleys form the two-component degenerate plasma and the low crystal symmetry leads to the strong anisotropy of its properties. The strong Fermi velocities contrast in different valleys provides the weakly coupled regime of the embodied two-component liquid, making the AP mode equivalent to the zero sound propagating in the slow component. The direct transformation between zero sound in weakly interacting case and the ion-acoustic wave can take place within a single crystal with the change of the wave propagation direction with respect to the crystallographic axes, as it was recently predicated for the case of type-I Weyl semimetals like TaAs~\cite{Afanasiev2021} by numerical calculations of the anisotropy of AP velocity. In addition to this, the smallness of the interaction constant averaged over the crystallographic directions ensures the absence of the renormalization effects due to the short-range interaction. Unlike shear sound, acoustic plasmon-zero sound mode in condensed matter systems is experimentally accessible by the full set of charge-sensitive experimental techniques, e.g. energy-loss spectroscopy and time-resolved measurements of the relaxation of electrostatic perturbations.

%%%%%%%%%%%%%%%%%%%%%%%%%%%%%%%%%%%%%%%%%%%%%%%%%%%%%%%%%%%%%%%%%%%%%%%%%%%%%%%%%%%%%%%%%%%%%%%%%%%%%%%%%%%
%%%%%%%%%%%%%%%%%%%%%%%%%%%%%%%%%%%%%%%%%%%%%%%%%%%%%%%%%%%%%%%%%%%%%%%%%%%%%%%%%%%%%%%%%%%%%%%%%%%%%%%%%%%
\section*{Acknowledgments}
The author is grateful to D.~Svintsov, A.~A.~Greshnov and G.~G.~Zegrya for valuable discussions and advice. This work was supported by the Foundation for the Advancement of Theoretical Physics and Mathematics BASIS (Grant No. 19-1-5-127-1).

%%%%%%%%%%%%%%%%%%%%%%%%%%%%%%%%%%%%%%%%%%%%%%%%%%%%%%%%%%%%%%%%%%%%%%%%%%%%%%%%%%%%%%%%%%%%%%%%%%%%%%%%%%%
%%%%%%%%%%%%%%%%%%%%%%%%%%%%%%%%%%%%%%%%%%%%%%%%%%%%%%%%%%%%%%%%%%%%%%%%%%%%%%%%%%%%%%%%%%%%%%%%%%%%%%%%%%%

\onecolumngrid

\appendix*

%%%%%%%%%%%%%%%%%%%%%%%%%%%%%%%
\section{Explicit form of the isotropic Landau parameters in three- and two-dimensional two-component electron liquids}
\label{Sec:Apdx}

At arbitrary values of Wigner-Seitz radii $r_{1,2}$, isotropic Landau parameters in two-component ELs can be calculated only numerically. The SX term $F_{\mathrm{SX}}^{\alpha\beta}$ is represented by the one-dimensional integral of~(\ref{Eq:LFsx}) over $\vartheta$. Angular integrations over directions of $\textbf{q}$ and $\textbf{k}'$ in the CH term $F_{\mathrm{CH}}^{\alpha\beta}$ can be done analytically and independently for the case of isotropic liquid, so that $F_{\mathrm{CH}}^{\alpha\beta}$ is given by the two-dimensional integral. Explicit expressions for the isotropic Landau parameters have the form

\begin{gather}
	\label{Apdx:F3sx}
	F_{0,\mathrm{SX}}^{\alpha \beta}=-\delta_{\alpha \beta} \frac{\gamma_3 r_{\alpha}}{\pi}\int\limits_0^1 \frac{u du}{u^2-\frac{\gamma_3 r_{\alpha}}{\pi}\sum\limits_{j=1,2}\frac{\kappa_j^2}{\kappa_{\alpha}^2}\tilde{\Pi}_3\left[\frac{k_{\alpha}}{k_j}u,0\right]} ,\\
	\label{Apdx:F3ch}
	F_{0,\mathrm{CH}}^{\alpha \beta}=-\frac{\gamma_3^2 r_{\alpha} r_{\beta}}{4\pi^3}\int\limits_0^{+\infty}du \int\limits_0^{+\infty}dw \frac{u \prod\limits_{j=\alpha,\beta} \mathcal{G}_3\left[\frac{k_{\beta}}{k_j}u,\frac{v_{\beta}}{v_j}w\right]}{\left(u^2-\frac{\gamma_3 r_{\beta}}{\pi}\sum\limits_{j=1,2}\frac{\kappa_j^2}{\kappa_{\beta}^2}\tilde{\Pi}_3\left[\frac{k_{\beta}}{k_j}u,i\frac{v_{\beta}}{v_j}w\right] \right)^2} ,
\end{gather}
in 3D and 
\begin{gather}
	\label{Apdx:F2sx}
	F_{0,\mathrm{SX}}^{\alpha \beta}=-\delta_{\alpha \beta} \frac{\gamma_2 r_{\alpha}}{\pi}\int\limits_0^1 \frac{du}{\sqrt{1-u^2}}\frac{1}{u-\gamma_2 r_{\alpha}\sum\limits_{j=1,2}\frac{\kappa_j}{\kappa_{\alpha}}\tilde{\Pi}_2\left[\frac{k_{\alpha}}{k_j}u,0\right]} ,\\
	\label{Apdx:F2ch}
	F_{0,\mathrm{CH}}^{\alpha \beta}=-\frac{\gamma_2^2 r_{\alpha} r_{\beta}}{\pi}\int\limits_0^{+\infty}du \int\limits_0^{+\infty}dw \frac{\prod\limits_{j=\alpha,\beta} \mathcal{G}_2\left[\frac{k_{\beta}}{k_j}u,\frac{v_{\beta}}{v_j}w\right]}{\left(u- \gamma_2 r_{\beta}\sum\limits_{j=1,2}\frac{\kappa_j}{\kappa_{\beta}}\tilde{\Pi}_2\left[\frac{k_{\beta}}{k_j}u,i\frac{v_{\beta}}{v_j}w\right] \right)^2} ,
\end{gather}
in 2D. Here $\gamma_3=(4/9\pi)^{1/3}$ and $\gamma_2=1/\sqrt{2}$. The dimensionless integration variables stand for $u=|\textbf{k}_{\alpha}-\textbf{k}_{\alpha}'|/2k_{\alpha}$ in the SX terms and for $u=q/2k_{\beta}$ and $w=\omega/v_{\beta}q$ in the CH ones. The functions $\mathcal{G}_d\left(\frac{k}{2 k_F},\frac{\omega}{v_F k}\right)=2^\frac{1-d}{4-d}\frac{k_F}{k}\mu^{-1}\sum\limits_{\Omega_{\bf k}}\left[G(\textbf{k}_F-\textbf{k},\mu-i\omega)+G(\textbf{k}_F+\textbf{k},\mu+i\omega)\right]$ given by
\begin{gather}
	\mathcal{G}_3(u,w)=\ln\frac{(1-u)^2+w^2}{(1+u)^2+w^2} ,\\
	\mathcal{G}_2(u,w)=-\left(\frac{\sqrt{(u^2-w^2-1)^2+4 u^2 w^2}+u^2-w^2-1}{(u^2 -w^2 -1)^2+4 u^2 w^2}\right)^{1/2} ,
\end{gather}
are proportional to the probability amplitudes of excitation of non-interacting fermion from the Fermi surface, averaged over the directions of the wave vector. Dimensionless RPA polarizabilities $\tilde{\Pi}_d(u,i w)$ at imaginary frequencies are real-valued functions even in $\omega$. They have the form
\begin{gather}
	\label{Apdx:P3}
	\tilde{\Pi}_3(u,iw)=-\frac{1}{2}+\frac{1-u^2+w^2}{8u}\ln\frac{(1-u)^2+w^2}{(1+u)^2+w^2}+\frac{w}{2}\left(\arctan \frac{1-u}{w}+\arctan\frac{1+u}{w}\right) ,\\
	\label{Apdx:P2}
	\tilde{\Pi}_2(u,iw)=-1+\frac{1}{\sqrt{2}u}\left(\sqrt{(u^2 - w^2 -1)^2+4 u^2 w^2}+u^2-w^2-1\right)^{1/2} .
\end{gather}
Note that $\tilde{\Pi}_d(u,iw)$ are strongly non-uniform in the $(u,w)$ space. Polarizabilities $\tilde{\Pi}_d(u,iw)\approx -1$ are finite for the electron transitions at the Fermi surface $0<u<1$, $w\ll 1$. Therefore, the corresponding coulomb interaction is substantially screened. However, the transitions outside the Fermi surface are affected by a rather weak dynamical screening, described by $\tilde{\Pi}_d(u,iw)\approx -1/d(u^2+w^2)$ at $u,w\gg 1$ .

\twocolumngrid

\bibliography{AP_LF_bib}

\begin{thebibliography}{64}
\bibitem [{Us_()}]{Us_Note}%
  \BibitemOpen
  \href@noop {} {\ }\bibinfo {note} {Here we assume that the amplitude of the
  self-consistent potential is real function.}\BibitemShut {Stop}%
\bibitem [{Ell()}]{Elliptic_Integrals_Note}%
  \BibitemOpen
  \href@noop {} {\ }\bibinfo {note} {In this work we use the following
  definitions of the complete elliptic
  integrals:~$K(m)=\int\limits_{0}^{\pi/2}(1-m \sin^2\theta)^{-1/2}d\theta$
  and~$E(m)=\int\limits_{0}^{\pi/2}(1-m
  \sin^2\theta)^{1/2}d\theta$.}\BibitemShut {Stop}%
\end{thebibliography}

\end{document}